\documentclass{article}

\usepackage{geometry,amsmath,amsfonts,algorithm,algorithmic,graphicx,subcaption,xcolor,nomencl,etoolbox,comment}

\title{\emph{CHIRRUP}:\\a practical algorithm for unsourced multiple access}
\author{Robert Calderbank and Andrew Thompson}

\begin{document}
\maketitle

\newcommand{\mC}{\mathcal{C}}
\newcommand{\CC}{\mathbb{C}}
\newcommand{\ZZ}{\mathbb{Z}}
\newcommand{\mO}{\mathcal{O}}
\newcommand{\RR}{\mathbb{R}}
\newcommand{\mH}{\mathcal{H}}
\newcommand{\PP}{\mathbb{P}}
\newcommand{\EE}{\mathbb{E}}
\newcommand{\PPP}{\mathcal{P}}

\newtheorem{theorem}{Theorem}
\newtheorem{corollary}{Corollary}
\newtheorem{proposition}{Proposition}

\newcommand{\edit}[1]{{ #1}}

\renewcommand{\nomname}{Glossary of notation}

\makenomenclature

\providetoggle{nomsort}
\settoggle{nomsort}{true} % true = sort by use, false = sort as usual
\makeatletter
\iftoggle{nomsort}{%
    \let\old@@@nomenclature=\@@@nomenclature        
        \newcounter{@nomcount} \setcounter{@nomcount}{0}%
        \renewcommand\the@nomcount{\two@digits{\value{@nomcount}}}% Ensure 10>01
        \def\@@@nomenclature[#1]#2#3{% Taken from package documentation
          \addtocounter{@nomcount}{1}%
        \def\@tempa{#2}\def\@tempb{#3}%
          \protected@write\@nomenclaturefile{}%
          {\string\nomenclatureentry{\the@nomcount\nom@verb\@tempa @[{\nom@verb\@tempa}]%
          \begingroup\nom@verb\@tempb\protect\nomeqref{\theequation}%
          |nompageref}{\thepage}}%
          \endgroup
          \@esphack}%
      }{}
\makeatother

\begin{abstract}
{Unsourced multiple access abstracts grantless simultaneous communication of a large number of devices (messages) each of which transmits (is transmitted) infrequently. It provides a model for machine-to-machine communication in the Internet of Things (IoT), including the special case of radio-frequency identification (RFID), as well as neighbor discovery in ad hoc wireless networks. This paper presents a fast algorithm for unsourced multiple access that scales to $\mC=2^{100}$ (active or non-active) devices (arbitrary $100$ bit messages). The primary building block is multiuser detection of binary chirps which are simply codewords in the second order Reed Muller code. The chirp detection algorithm originally presented by Howard et al.~\cite{chirp_reconstruction} is enhanced and integrated into a peeling decoder designed for a patching and slotting framework. In terms of both energy per bit and number of active devices (number of transmitted messages), the proposed algorithm is within a factor of $2$ of state of the art approaches. A significant advantage of our algorithm is its computational efficiency. We prove that the worst-case complexity of the basic chirp reconstruction algorithm is $\mO[nK(\log_2^2 n + K)]$, where $n$ is the codeword length and $K$ is the number of active users. Crucially the complexity is sublinear in $\mC$, which makes the reconstruction computationally feasible --- a claim we support by reporting computing times for our algorithm. Our performance and computing time results represent a benchmark against which other practical algorithms can be measured.}
{Unsourced multiple access, chirp codes, compressed sensing, reconstruction algorithms.}
\end{abstract}

\newpage

\nomenclature{$B$}{Number of bits per message.}
\nomenclature{$\mC$}{Total number of devices/messages, equal to $2^B$.}
\nomenclature{$K$}{Number of active devices or transmitted messages.}
\nomenclature{$s$}{Activity pattern to recover.}
\nomenclature{$x_i$}{Codeword corresponding to message $i$.}
\nomenclature{$X$}{Codebook matrix whose columns are $\{x_i\}$.}
\nomenclature{$z$}{Additive noise.}
\nomenclature{$y$}{Measurement vector.}
\nomenclature{$\left\{\phi_{P,b}\right\}$}{Binary chirp with matrix $P$ and translate $b$.}
\nomenclature{$P$}{$m\times m$ binary symmetric matrix.}
\nomenclature{$b$}{Translate in $\ZZ_2^m$.}
\nomenclature{$Q$}{Transmitted power per message.}
\nomenclature{$\mH(\cdot)$}{Walsh-Hadamard transform.}
\nomenclature{$c$}{Number of children in the tree search enhancement of CHIRRUP.}
\nomenclature{$2^p$}{Number of slots.}
\nomenclature{$2^r$}{Number of patches.}
\nomenclature{$\tilde{n}$}{Codeword length in each slot, equal to $2^m$ for some $m$.}
\nomenclature{$n$}{Overall codeword length, equal to $2^{m+p+r}$.}
\nomenclature{$n$}{Codeword length, equal to $2^m$ for some $m$.}
\nomenclature{$l$}{Vector giving number of parity check bits appended to each patch.}
\nomenclature{$E_i$}{Error event associated with message $i$.}
\nomenclature{$\PPP$}{Per-user probability of error.}
\nomenclature{$\epsilon$}{Error threshold.}
\nomenclature{$\Lambda$}{Set of transmitted messages.}
\nomenclature{$\Gamma$}{Set of recovered messages.}

\printnomenclature

\newpage

\section{Introduction}\label{intro}

One of the opportunities that justifies the development of a new generation of wireless networks (5G) is machine-to-machine (M2M) communication (see~\cite{5G}). This Internet of Things (IoT) will support grantless, simultaneous communication of a large network of devices, each of which can be thought to transmit infrequently. In this paradigm the messages are small and the number of users is massive, and it is the messages rather than the identities of the users that must be recovered. The development of next generation wireless networks is guided by information theory which provides performance bounds that govern transmission of short packets (see~\cite{IoT}).

\subsection{Unsourced multiple access}

We follow the framework proposed by Polyanskiy \cite{polyanskiy} in which communication occurs in blocks of $n$ channel uses (measurements) and the the task of a receiver is to correctly identify $K$ messages, each consisting of $B$ bits. A typical regime of interest is $n=30000$, $K=300$ and $B=100$. As in~\cite{polyanskiy}, we model this scenario using a Gaussian multiple access channel in which the receiver observes
$$y=\sum_{i=1}^{\mC}s_i x_i + z,$$
where $\mC:=2^B$ is the total number of possible $B$-bit messages, $s=\begin{bmatrix}s_1&\dots&s_{\mC}\end{bmatrix}^T\in\{0,1\}^{\mC}$ is the activity pattern with Hamming weight $K$, $x_i\in\RR^n$ is the codeword corresponding to message $i$ and $z\sim\mathcal{N}(0,I_n)$ is additive white Gaussian noise (AWGN). Writing $X=\begin{bmatrix}x_1&\dots&x_{\mC}\end{bmatrix}$, we may represent the encoding as the matrix equation 
$$y=Xs+z.$$

One popular approach to the problem is slotted ALOHA~\cite{slotted_ALOHA} in which a frame is divided into subframes and each message is sent over a randomly selected subframe (or slot). Decoding of a given message is successful if it encounters no collision with other messages. This no-collision requirement imposes a severe constraint on the number of messages that can be sent, which can be partially alleviated using Coded Slotted ALOHA~\cite{coded_ALOHA1,coded_ALOHA2,coded_ALOHA3} in which messages are sent to multiple (two or more) slots and information is passed between slots to recover messages originally lost to colllisions. These contributions assume synchronized transmission and perfect interference cancellation. The asynchronous model is considered in~\cite{asynchronous} under the assumption of perfect interference cancellation.

The point is made in~\cite{ordentlich} that slotted ALOHA only supports single user detection within each slot, and the authors instead propose $T$-fold ALOHA in which up to $T$ messages can be jointly decoded in each slot. Within each slot, messages are encoded using a product of two codes: the inner code enables the receiver to decode the modulo-$2$ sum of all codewords transmitted and the outer code allows the receiver to recover the individual messages from the sum. Rather than proposing a specific algorithm, the authors obtain theoretical bounds on performance using fundamental coding limits for the inner code. These bounds are shown to give an improvement upon standard slotted ALOHA-based methods in terms of energy per bit. In practice, a suitable coding and decoding scheme for the binary memoryless channel must be chosen (for example LDPC codes and message-passing recovery), and the choice of decoding algorithm involves a tradeoff between computational efficiency and suboptimality with respect to the coding limits. 

Further bounds are given in~\cite{replica} in which the replica symmetric formula is used to analyze the achievable rate of approximate message passing for decoding the inner code, and improved bounds are reported. The unsourced multiple access paradigm was further extended to the case of massive MIMO and multiple antennas in~\cite{massive_MIMO}.

In~\cite{krishna}, a method was proposed which also performs joint decoding within a slotted framework, but with a number of differences to the approach in~\cite{ordentlich}. Rather than assigning messages to slots at random, the slots in which a message is repeated are a deterministic function of the message itself. Within each slot, part of the message is encoded by an LDPC code. A preamble picks an interleaver (permutation) for the LDPC code, and is encoded using either a random Bernoulli or a BCH codebook. Either one-step thresholding (treat interference as noise) or nonnegative least squares is used to decode the preamble, and this information is then fed into a message passing algorithm for decoding the interleaved LDPC code. Information is also allowed to pass between slots: when a message is decoded, its corresponding codewords are `peeled off' all slots within which it was transmitted. The authors demonstrate further energy per bit improvements over~\cite{ordentlich}.

In~\cite{krishna_CS}, a divide-and-conquer compressed sensing approach is taken in which the message is encoded with redundancy in the form of parity check constraints determined by a linear block code. The message is then split into \emph{patches}, each of which is encoded using a BCH codebook. Each patch is reconstructed using nonnegative least squares, and the messages are then stitched together using a tree decoder. The full paper~\cite{krishna_CS_full} analyzes two different parity bit allocation strategies in the limit as the number of users and their corresponding payloads tend to infinity. The number of channel uses needed and the computational complexity associated with these allocation strategies are explicitly characterized for various scaling regimes. Further improvements in terms of energy per bit are reported, though for a more modest number of bits than in~\cite{ordentlich,krishna}. 

\subsection{Main contributions}

Chirps are a broad class of signals with optimal auto- and cross-correlation properties that encourage their use in multiuser sonar and radar, and in communications (for references to these applications see~\cite{brodzik,bat_chirps}). 

We propose CHIRRUP,\footnote{\texttt{MATLAB} code is available to download from \texttt{https://github.com/ajthompson42/CHIRRUP}.} a novel compressed sensing algorithm for unsourced multiple access which employs \emph{binary} chirp coding and the chirp reconstruction algorithm originally proposed in~\cite{chirp_reconstruction}. More specifically, we enhance the original algorithm in a number of ways which are inspired by the aforementioned prior work. In particular:
\begin{enumerate}
\item Following~\cite{slotted_ALOHA,coded_ALOHA1,coded_ALOHA2,ordentlich,krishna}, we incorporate the algorithm into a slotting framework in which each message is sent to precisely two slots. It is this repeated transmission of chirps which inspires the name of the algorithm. \edit{As in~\cite{krishna}, the slotting pattern is a deterministic function of the message bits, which allows information to be encoded in the slotting pattern.} We design a coding scheme which allows messages to be passed between slots: Each messages is assigned two slots, and when a message is decoded in one slot the corresponding codewords in its twin slot is peeled off, in a similar spirit to~\cite{krishna}. Our algorithm cycles through the slots allowing the decoded results to propagate.
\item We build in the option to split the message into a small number of patches with the addition of extra parity check constraints. Each patch is decoded separately, and the tree decoder proposed in~\cite{krishna_CS} is then employed to patch together the full message.
\end{enumerate}

One significant advantage of our approach is computational efficiency. One additional contribution of this paper is to analyse the complexity of the chirp reconstruction algorithm, and show that its worst-case complexity is $\mO[nK(K+\log_2^2 n)]$, where $n$ is the codebook length and $K$ is the number of transmitted messages. In contrast to almost all algorithms for compressed sensing, our enhanced chirp reconstruction algorithm is sublinear in the number of codewords, which makes it an attractive choice for a problem whose scale is massive compared to those that are typically considered. We present numerical results which show that the performance of our algorithm in terms of energy per bit is comparable with the results presented in~\cite{ordentlich,krishna,krishna_CS} and the maximum number of messages that can be recovered is within a factor of $2$ of the results in~\cite{ordentlich,krishna,krishna_CS}. Meanwhile we present for the first time in this context information on practical computational efficiency in the form of running time results. Our results therefore represent a benchmark against which other practical algorithms for unsourced multiple access may be compared.

The structure of the rest of the paper is as follows. In Section~\ref{basic}, we describe the original chirp reconstruction algorithm proposed in~\cite{chirp_reconstruction} and give a worst-case complexity result. In Section~\ref{enhancements}, we describe in detail the enhancements to the original algorithm which lead to improved performance on the unsourced multiple access problem. In Section~\ref{experiments}, we present the results of our numerical experimentation, before offering conclusions and reflections on future work in Section~\ref{conclusions}.

\section{The chirp reconstruction algorithm}\label{basic}

Our basic approach is to assign to each message a codeword from a codebook of binary chirps. Assuming that $n=2^m$ for some $m$, we index the $2^m$ entries by a binary $m$-tuple $v\in\ZZ_2^m$ and consider \emph{binary chirp} codewords of the form $\{\phi_{P,b}\}$ where
\begin{equation}\label{chirp_def}
\{\phi_{P,b}\}_v:=\sqrt{Q}\cdot i^{2b^T v+v^T Pv},
\end{equation}
where addition in the exponent is modulo $4$.\footnote{All quantities in the exponent are defined over $\ZZ_2$, but the exponent itself is computed over $\ZZ_4$, for example
$$2\begin{bmatrix}1&0&0\end{bmatrix}\begin{bmatrix}1\\1\\1\end{bmatrix}+\begin{bmatrix}1&1&1\end{bmatrix}\begin{bmatrix}1&1&0\\1&0&0\\0&0&0\end{bmatrix}\begin{bmatrix}1\\1\\1\end{bmatrix}=2+3\equiv 1\;\;\textrm{mod}\;4.$$} The codebook consists of all $\{\phi_{P,b}\}$ where $b\in\ZZ_2^m$ is a binary $m$-tuple and $P\in\ZZ_2^{m\times m}$ is an $m\times m$ binary symmetric matrix. A binary $m$-tuple $b$ is uniquely determined by $m$ bits, while an $m\times m$ binary symmetric matrix $P$ is determined by $\frac{1}{2} m(m+1)$ bits, which means that the full codebook encodes $m+\frac{1}{2}m(m+1)=\frac{1}{2} m(m+3)$ bits. Note that, since $P$ is symmetric, 
$$v^T Pv=\sum_{i,j}v_i P_{ij} v_j=2\sum_{i<j}v_i P_{ij} v_j+\sum_i v_i P_{ii}.$$
If a real codebook is desired, we use only the subset of $m\times m$ binary symmetric matrices with zero diagonal, and in this case the codebook encodes slightly fewer bits, namely $m+\frac{1}{2}m(m-1)=\frac{1}{2} m(m+1)$. In either case, note that the number of bits is quadratic in $m=\log_2 n$. Each codeword satisfies $\|\phi_{P,b}\|_2^2=nQ$, and so the constant $Q$ in (\ref{chirp_def}) represents the transmitted power per message. Given $K$ active messages, our measurements therefore take the form
\begin{equation}\label{measurements}
y=\sum_{l=1}^K \phi_{P_l,b_l}+z,
\end{equation}
where $z\sim\mathcal{N}(0,I_n)$.

The main building block of our decoding strategy is the \emph{chirp reconstruction algorithm} originally proposed by Howard et al. \cite{chirp_reconstruction}. The complexity of this algorithm is sublinear in the number of codewords, a vital consideration in the problem at hand if the number of codewords is of the order $2^{100}$. We will show that the worst-case complexity of the algorithm is $\mO\big[nK(K+\log_2^2 n)\big]$, where $K$ is the number of messages. The algorithm is greedy in nature, finding in each iteration a single active component (codeword) and performing a least-squares fit over all components so far discovered, and in that sense it is similar to the popular Orthogonal Matching Pursuit (OMP) algorithm. Where it differs from OMP is in the procedure for identifying the new component. OMP calculates the correlation between the current measurement residual and each column and selects the column whose correlation has the largest absolute value. Finding the maximum value from a list of $\mC$ candidates in this way is an $\mO(\mC)$ procedure and therefore computationally intractable for the problem at hand. By contrast, the chirp reconstruction algorithm performs a small number of $\mO(n\log_2 n)$ transforms to identify the matrix $P$ row-by-row, and subsequently to identify the vector $b$, corresponding to an active codeword. The chirp reconstruction algorithm is summarized in Algorithm~\ref{chirp_alg}. The iteration limit $S$ should be chosen to be a small multiple of the number of messages.\footnote{In practice we take $S$ to be three times the expected number of messages.}

\begin{algorithm}
\textbf{Inputs}: Measurements $y\in\RR^n$ where $n=2^m$; target sparsity $S$.\\
\textbf{Initializations:} $\Phi=\begin{bmatrix}\;\end{bmatrix}$; iteration count $s=0$.
\begin{algorithmic}
\WHILE {$\|y\|>\epsilon$ and $s\le S$}
\STATE $s\leftarrow s+1$
\STATE $(P_s,b_s)\leftarrow\mathbf{findPb}(y)$
\STATE $\Phi\leftarrow\begin{bmatrix}\Phi &\phi_{P_s,b_s}\end{bmatrix}$
\STATE $c_s\leftarrow\Phi^{\dag} y$
\STATE $y\leftarrow y - \Phi c_s$
\ENDWHILE
\end{algorithmic}
\textbf{Output:} $\{P_s,b_s,c_s\}$.
\caption{Chirp Reconstruction Algorithm~\cite{chirp_reconstruction}}
\label{chirp_alg}
\end{algorithm}

\begin{figure}[t!]
\centering
\begin{subfigure}[b]{0.45\textwidth}{\includegraphics[width=\textwidth]{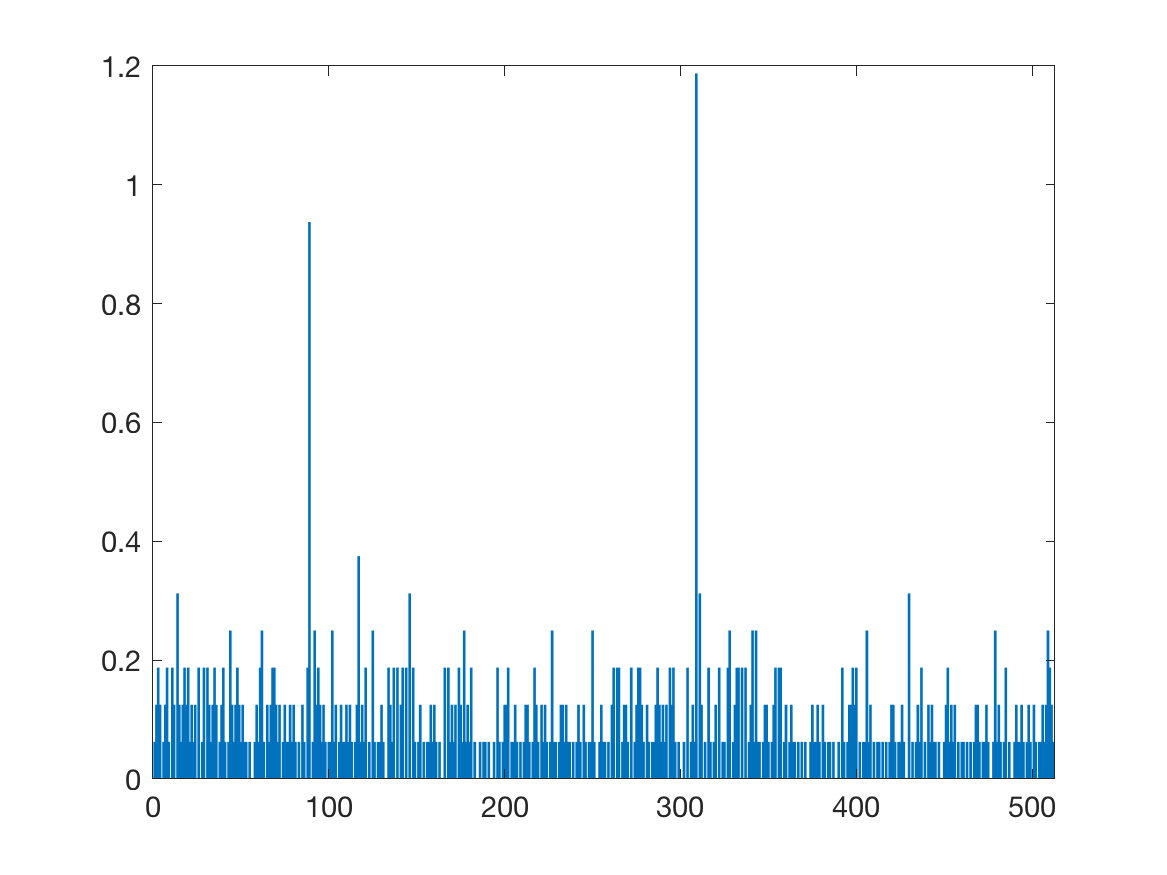}\caption{$e:=e_2$.}}\end{subfigure}
\begin{subfigure}[b]{0.45\textwidth}{\includegraphics[width=\textwidth]{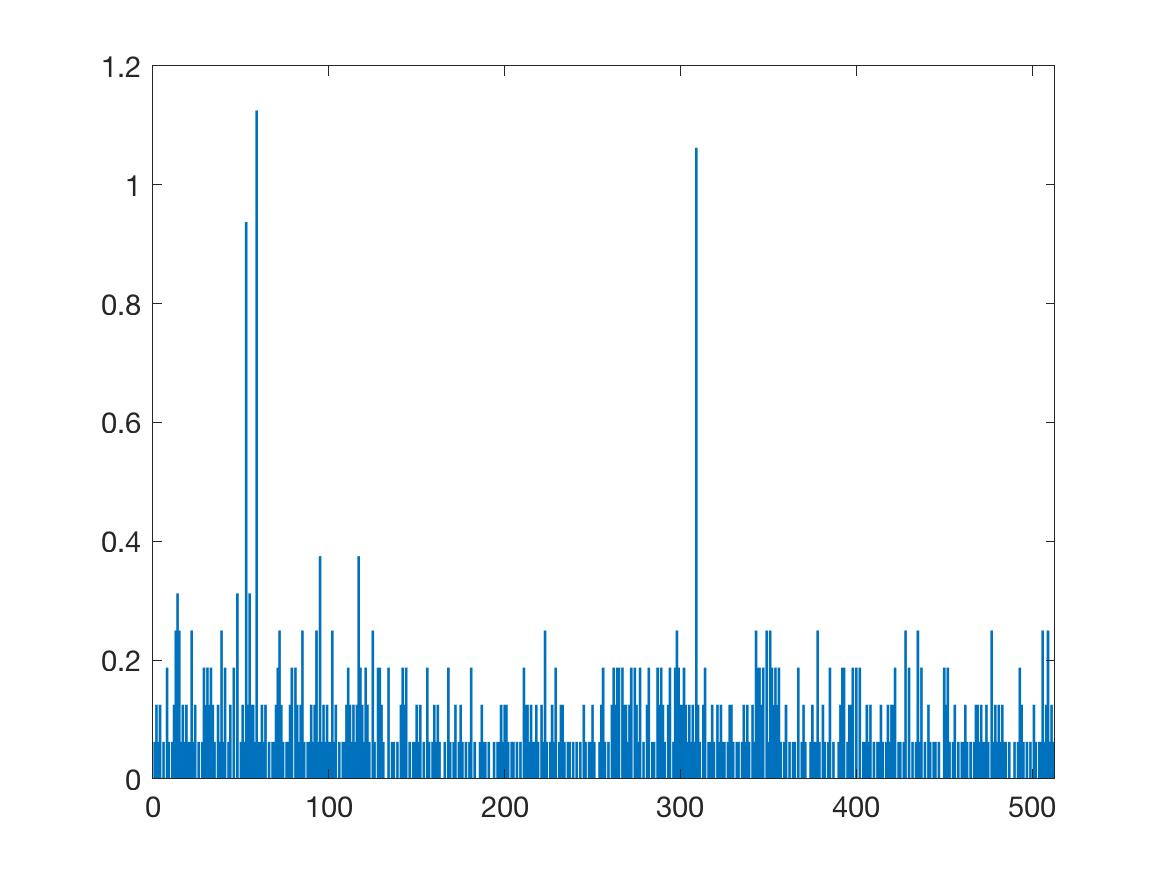}\caption{$e:=e_2+e_1$.}}\end{subfigure}
\begin{subfigure}[b]{0.45\textwidth}{\includegraphics[width=\textwidth]{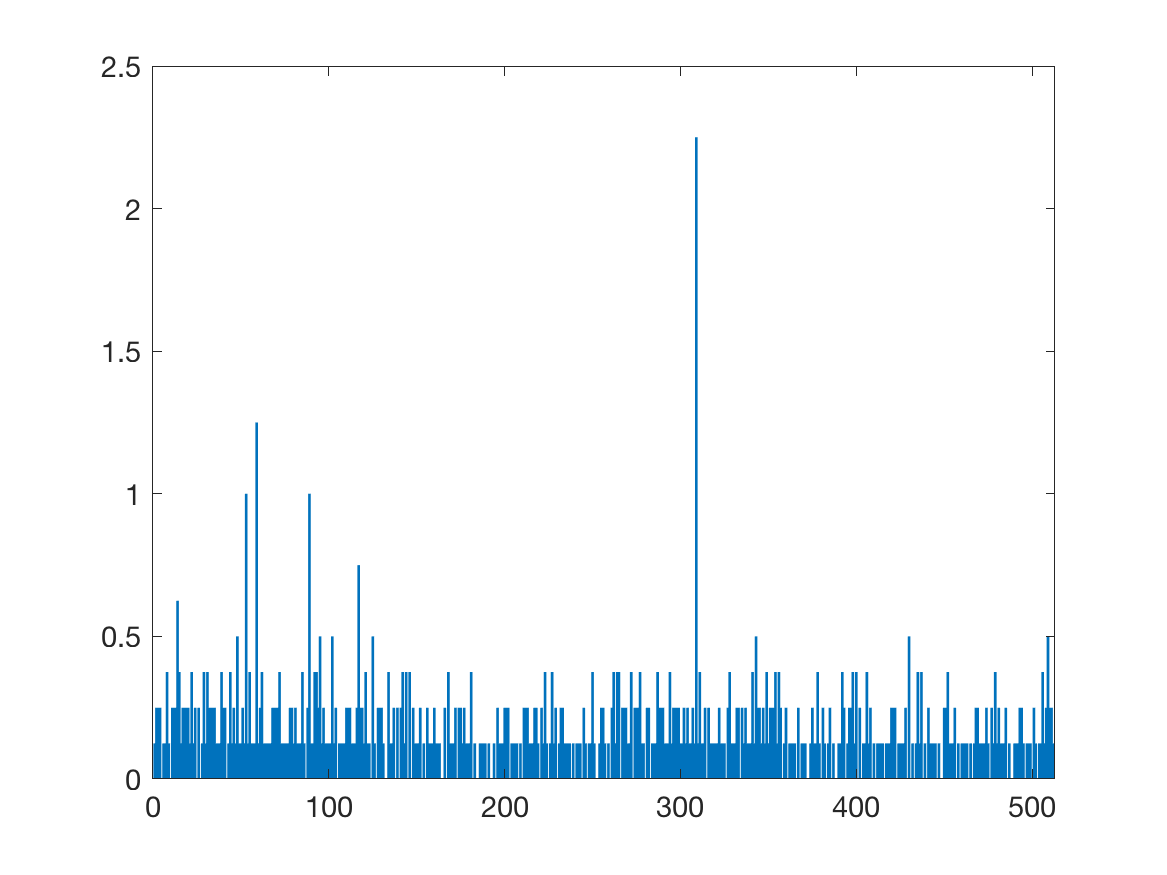}\caption{The sum of the two absolute values.}}\end{subfigure}
\caption{A measurement consisting of the noiseless superposition of $4$ chirp components ($m=10$) is generated, and the figure plots the absolute values of $f^e$ in the determination of the second row of the $P$ matrix in the first iteration. Out of the $512$ remaining candidates, a small number emerge above the noise floor. When the absolute $f^e$ scores for $e:=e_2$ and $e:=e_2+e_1$ are summed, a single component emerges, which corresponds to the correct second row.}\label{noise_floor}
\end{figure}

$\mathbf{findPb}$ denotes the subroutine for identifying a new component, which is based upon the application of shift-and-multiply to the measurements. More specifically, we shift the binary indices of $y$ through addition of some $e\in\ZZ_2^m\setminus{0}$, and then multiple componentwise with the complex conjugate of $y$ to obtain a vector $f^e\in\ZZ_2^m$ whose components are $f^e_a:=\overline{y_a}y_{a+e}$. Writing $y_a$ for the component of $y$ indexed by $a\in\ZZ_2^m$, we have the following expression for $f^e$.
\begin{proposition}\label{shift_multiply}
For any $e\in\ZZ_2^m$, the following holds for all $a\in\ZZ_2^m$.
\begin{eqnarray}\label{peaks}
f^e_a&=&\sqrt{Q}\sum_{k=1}^K \left\{\phi_{P_k,b_k}\right\}_e(-1)^{e^T P_k a}+\sum_{k\neq l}\left\{\phi_{P_l,b_l}\right\}_e\left\{\phi_{P_l-P_k,b_k-b_l+P_l e}\right\}_a\nonumber\\&& \;\;\;\;+\overline{z}_a \left[\sum_{k=1}^K \phi_{P_k,b_k}\right]_{a+e}+z_{a+e}\overline{\left[\sum_{k=1}^K \phi_{P_k,b_k}\right]_a}+\overline{z}_a z_{a+e}.
\end{eqnarray}
\end{proposition}
A proof of Proposition~\ref{shift_multiply} can be found in Appendix~\ref{shift_multiply_proof}. The right-hand side of (\ref{peaks}) is a linear combination of Walsh functions (the first term) plus a linear combination of chirps (the remaining terms). We may think of chirps as being distributed across all Walsh functions to equal degree, and therefore these cross terms appear as a uniform noise floor. The Walsh-Hadamard transform $\mH(z)$ is the decomposition of a signal $z\in\ZZ_2^m$ into a basis of Walsh functions, and given $v\in\ZZ_2^m$ we define its coefficients in the usual way as
$$\left\{\mH(z)\right\}_v:=\frac{1}{2^{m/2}}\sum_{a\in\ZZ_2^m}(-1)^{v^T a}z_a.$$
The Walsh-Hadamard Transform is the binary analogue of the Discrete Fourier Transform, and applying the Walsh-Hadamard transform $\mH$ to (\ref{peaks}) results in peaks at `frequencies' $P_k e$ for each $P_k$ in the data. Useful information can thus be obtained by choosing any $e\in\ZZ_2^m\setminus 0$. In particular, if $e:=e_r$ is the $r$th canonical basis vector in $\ZZ_2^m$, peaks will correspond to the $r$th row (or column) of the $\{P_k\}$ matrices. In this way, an active $P_k$ matrix can be discovered row-by-row. To mitigate cancellation effects for $r>1$, a second choice of $e:=e_r+e_{r-1}$ is also used, the absolute values of the two outputs are summed, and the peak then chosen (see~\cite{chirp_reconstruction} for further explanation). By way of illustration, Figure~\ref{noise_floor} plots the absolute values of $f$ in the determination of the second column of the $P$ matrix of a component. The figure illustrates how components corresponding to chirps present in the signal are prominent above the `noise floor'. Having found $P_k$, the corresponding $b_k$ can be determined by \emph{dechirping} the signal by calculating $i^{-a^T P_k a}y_a$ for each $a\in\ZZ_2^m$ and applying the Walsh-Hadamard transform $\mH$ to the result. The $\mathbf{findPb}$ algorithm is summarized in Algorithm~\ref{findPb}.

%\footnote{In reality, we consider the first $Q$ candidates for the rows of $P_l$ with the largest peaks in descending order of amplitude, terminating and accepting the current choice if the peak corresponding to $b_l$ in the dechirped signal is a factor $\beta$ above the average RMSE of the other components. Sensible choices for the parameters $Q$ and $\beta$ are $Q=3$ and $\beta=3$.}

\begin{algorithm}
\textbf{Inputs:} Residual $y\in\RR^n$.\\
\textbf{Initializations:} $P=\begin{bmatrix}\;\end{bmatrix}$.
\begin{algorithmic}
\FOR {$r=1$ to $m$}
\STATE $f_a\leftarrow \overline{y_a}y_{a+e_r}$ for $a\in\ZZ_2^m$
\IF {r=1}
\STATE $h\leftarrow|\mH(f)|$
\ELSE
\STATE $g_a\leftarrow \overline{y_a}y_{a+e_r+e_{r-1}}$ for $a\in\ZZ_2^m$
\STATE $h\leftarrow|\mH(f)|+|\mH(g)|$
\ENDIF
\STATE $p_r\leftarrow$ location of largest entry of $h$
\ENDFOR
\STATE $P\leftarrow\begin{bmatrix}p_1&\ldots&p_m\end{bmatrix}$
\STATE $z_a\leftarrow i^{-a^T P a}y_a$ for $a\in\ZZ_2^m$
\STATE $w\leftarrow|\mH(z)|$
\STATE $b\leftarrow$ location of largest entry of $w$
\end{algorithmic}
\textbf{Outputs:} $P$, $b$.
\caption{\textbf{findPb}~\cite{chirp_reconstruction}}
\label{findPb}
\end{algorithm}

The most expensive operations in each iteration of Algorithm~\ref{findPb} are the Walsh-Hadamard transforms (WHTs), which can be computed in $\mO(n\log_2 n)$ operations. Two WHTs must be computed in each iteration to select $P$, and then a single WHT is required to select $b$. Since there are $m=\log_2 n$ iterations, the overall complexity of Algorithm~\ref{findPb} is therefore $\mO(n\log_2^2 n)$. As well as a call to Algorithm~\ref{findPb}, each iteration of Algorithm~\ref{chirp_alg} also requires the updating of the least-squares fit to the residual. Rather than performing these calculations explicitly, a more computationally efficient approach is the one given in~\cite{gradient_pursuits}, in which a reduced QR factorization is updated, which requires $\mO(nK)$ operations per iteration. Since there are $\mO(K)$ iterations in Algorithm~\ref{chirp_alg}, its overall complexity is therefore $\mO[nK(K+\log_2^2 n)]$. Note that, since $n\ll\mC$, the algorithm is massively sublinear in $\mC$.

We also consider an enhancement to Algorithm~\ref{findPb} which gives an improvement in performance at the cost of a higher worst-case complexity. The change is to allow the algorithm to test a number of candidates $c$ for each row of the matrix $P$ corresponding to the largest $c$ entries of $h$ (see Algorithm~\ref{findPb}). The choice of $P$ corresponds to selecting a branch of a tree in which the $r$th level of the tree represents the choice of the $r$th row of $P$, and in which each node has $c$ children. The tree is explored in a depth-first fashion, starting with the candidates at each level for which $h$ is greatest as in Algorithm~\ref{findPb}. When a branch of the tree is reached, the choice of $(P,b)$ is accepted if the absolute value of $w$ (see Algorithm~\ref{findPb}) exceeds some multiple $\alpha$ of the root mean squared error of $w$ restricted to the other components: we find $\alpha:=3$ to be a sensible choice. If not, we retrace our steps back along the branch of the tree, trying the other candidates for the $m$th row, followed by the $(m-1)$th row, and so on. If the whole tree is explored and no solution is found that satisfies the acceptance criterion, a random $(P,b)$ is generated.

Analyzing the worst-case complexity of this enhanced Algorithm~\ref{findPb}, we see that, if the entire tree is explored, $2\times c^r$ WHTs must be computed at level $r$ of the tree to select $P$, the sum total of which may be bounded above by $2c^{m+1}$, and $c^m$ WHTs must be computed at each of the branches. The total number of WHT computations is therefore bounded above by $3c^{m+1}=3cn^{\log_2 c}$. The worst-case complexity of the enhanced Algorithm~\ref{findPb} is therefore $\mO(n^{1+\log_2 c}\log_2 n)$. For example, we have $\mO(n^2\log_2 n)$ when $c=2$ and $\mO(n^3\log_2 n)$ when $c=4$. We find that $c=3$ is quite sufficient in practice. We also observe that in practice running times only increase by a linear factor compared to the basic algorithm since it is rare for the entire tree to be searched. The worst-case complexity of the full chirp reconstruction algorithm, Algorithm~\ref{chirp_alg}, with the enhancement, is therefore $\mO[nK(K+n^{1+\log_2 c}\log_2 n)]$, which is still sublinear in $\mC$ since $c\ll n$.

\section{The CHIRRUP algorithm}\label{enhancements}

The number of messages that can be successfully decoding using the basic algorithm scales somewhat poorly with codeword length. We thus consider three enhancements to the basic decoder: \emph{slotting}, \emph{message passing} and \emph{bit partitioning}. In this section, we assume the use of complex binary chirps, but each of the enhancements can also be used in the case of real binary chirps. In Section~\ref{experiments}, we will present numerical results using both complex and real binary chirps. We refer to the algorithm obtained by incorporating each of these enhancements into Algorithm~\ref{chirp_alg} as CHIRRUP.

\subsection{Slotting}

An enhancement which allows a greater number of messages to be recovered is slotting, in which a codeword length $n=2^{m+p}$ is divided into $2^p$ slots, with each message being sent to one or more of the slots. Equivalently, each message is assigned a sparse codeword which is zero except for the blocks to which the message is sent. The chirp reconstruction algorithm is then used to identify the messages within each slot, and the results are combined. Slotting can also be leveraged to send information by allowing some of the message bits to encode the slot location.

Slotting is frequently used in multiple access as a tool to manage interference, for example in slotted ALOHA protocols~\cite{slotted_ALOHA}. Slotting schemes have recently been proposed in the context of unsourced multiple access in~\cite{polyanskiy,ordentlich,krishna}. If the same `inner' code is used in each slot, this approach can also be viewed as a special case of a product code where the `outer' codebook matrix is simply the identity matrix.

Since slots must be assigned to messages \emph{a priori}, messages must be slotted in an uncoordinated fashion, which means that it is not possible to ensure a precise distribution of messages between the slots. However, if the slots are equally distributed between the messages and if the transmitted messages are independent of this distribution, one expects the number of active messages over the slots to follow a multinomial distribution with equal probabilities per slot.

\begin{comment}Using a slotted setup, we transmit $\frac{1}{2}m(m+3)+p$ bits by using $\frac{1}{2}m(m+3)$ bits to assign to each message a codeword from the codebook of all length-$2^m$ binary chirps, and by using the remaining $p$ bits to encode the location of the slot using an arbitrary subset of $p$ of these bits.
\end{comment}

\subsection{Message passing and a peeling decoder}\label{message}

Message passing algorithms are often used for solving inference problems on graphs, and in the context of wireless communication belief propagation decoders are frequently used for codes built from sparse graphs, such as LDPC codes and Turbo codes. For example in LDPC codes, the individual parity checks that comprise the code are decoded separately and the information shared iteratively. Exporting the principle to slotted chirp reconstruction, one might expect it to be advantageous to assign each message to multiple slots and to allow the slots to propagate the decoded messages in some iterative fashion. 

Our approach in general is as follows. The bits used to encode $(P,b)$ are given a dual function, namely also to encode the $q$ slots to which each message is sent. The decoding algorithm makes a number of passes (cycles) through all the slots, and whenever a $(P,b)$ pair is found, the slots to which it was sent can be identified and the corresponding chirp component `peeled off' from each set of measurements. In this fashion, the information decoded in each slot is propagated.

We found that sending each message to \emph{exactly two} ($q=2$) of the $2^p$ slots was optimal in practice. In this context, we also found a way to engineer the encoding of additional bits in the slot pattern, as follows. We transmit $\frac{1}{2}m(m+3)+p-1$ bits by assigning each message to $q=2$ out of the $2^p$ slots. Assuming $p\le\frac{1}{2}m(m+3)-1$, we append $p$ bits to encode a \emph{primary slot location}. From the bits used to encode $(P,b)$, we use an arbitrary subset of size $p$ to encode a \emph{translate}. By a translate, we mean that it gives the secondary slot location when it is added (using binary addition) to the primary slot location.\footnote{Specifically, we first use bits used to encode $b$, and if needed some of the bits used to encode $P$.} To allow the primary and secondary slots to be distinguished, we fix a single check digit in the $P$ matrix to be $0$ for the primary slot and $1$ for the secondary slot, deducting $1$ from the total number of bits transmitted. 

Our decoding algorithm makes $d$ cycles through all the slots, in each of which all of the slots are separately decoded.\footnote{We find $d=5$ to be universally sufficient in practice.} Whenever a $(P,b)$ pair is found, the check digit in the $P$ matrix reveals whether the current slot is primary or secondary for the corresponding message. Meanwhile the translate implicitly encoded in the given $(P,b)$ determines the other slot to which the message was sent. Regardless of whether the slot is primary or secondary, the primary slot is now known and the corresponding message can be decoded. Provided its corresponding coefficient $c$ is suitably close to $1$,\footnote{Various criteria are possible here and our choices are \emph{ad hoc}. For complex chirps, we found the condition $|c-1|<0.3$ to work well. For real chirps, we require $|\textrm{Re}(c)-1|<0.1$ and $|\textrm{Im}(c)|<0.1$.} the check digit is flipped and the resulting $(P,b)$ is passed to the other slot. Each time decoding is attempted in a given slot, a list of $(P,b)$ already known to be active in that slot (either from its own previous decodings or from messages passed from other slots) is compiled. The corresponding components are then `peeled off' from the signal before the iterative procedure begins, allowing more components to be found.

\subsection{Bit partitioning}

We have already noted that a given binary chirp codebook places an upper limit on the length of messages that can be transmitted, and the use of slotting further constrains the length of messages. Indeed a tradeoff can be observed at the outset here: for a fixed codebook length $n=2^m$, more slots allow more messages to be transmitted while reducing the length of messages that can be transmitted. This tradeoff will be explored quantitatively in Section~\ref{experiments}. One way to allow longer messages to be transmitted is to partition the message into patches in such a way that the patches can be decoded separately and then `patched together' after the event. In this regard, we adopt the approach recently proposed in~\cite{krishna_CS} in which each patch is encoded using a linear block code which introduces redundancy in the form of random parity checks. The patches are then decoded separately and `patched together' after the event using a tree encoder. We refer the reader to~\cite{krishna_CS} for further details of the method.

\begin{figure}
\centering
\includegraphics[width=\textwidth]{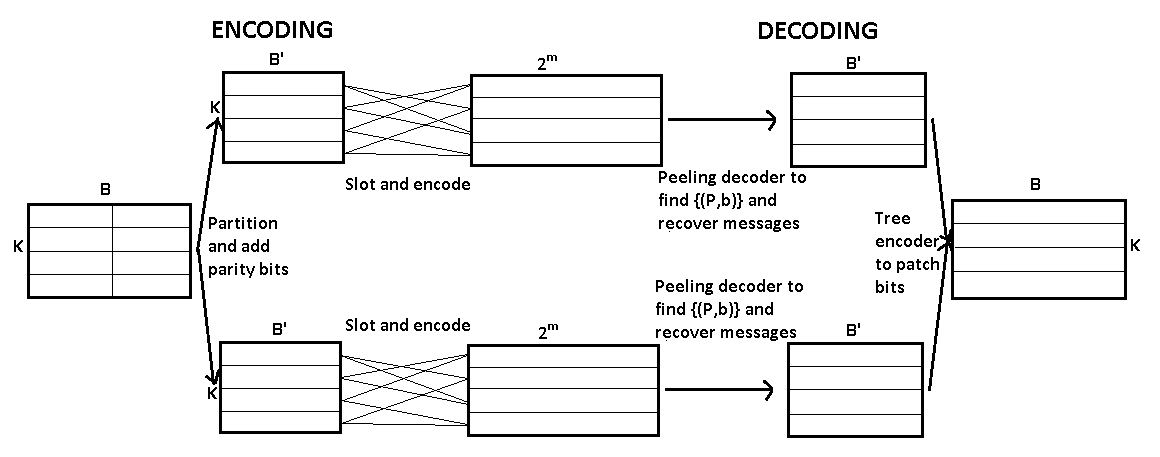}
\caption{An illustration of the slotting and patching schemes used in the CHIRRUP algorithm. For simplicity, we here illustrate for $2$ patches and $K=4$.}
\label{slot_patch}
\end{figure}

In our case, we choose an integer partition parameter $r$ and partition into $2^r$ patches ($r=0$ is a single patch, $r=1$ means two patches, etc.). We partition the length-$n=2^{m+p+r}$ codeword into $2^r$ sub-blocks, each of length $2^{m+p}$. We then further split each sub-block into $2^p$ slots of length $2^m$, as described in Section~\ref{message}. In this way a message consisting of $\frac{1}{2}m(m+3)+p-1$ bits can be sent patchwise in each of the sub-blocks. In keeping with~\cite{krishna_CS}, we append $l_i$ random parity check bits to patch $i$, for $i=2,\ldots,2^r$ (we never assign parity check bits to the first patch). These parity check bits are needed to patch together the messages, but do not serve the purpose of transmitting information.\footnote{The choice of $\{l_i\}$ depends upon the number of messages being transmitted (see~\cite{krishna_CS} for an illuminating theoretical analysis).} The number of bits of information transmitted is therefore
$$B=2^r\left\{\frac{1}{2} m(m+3)+p-1\right\}-\sum_{i=2}^r l_i.$$
An illustration of the slotting and patching scheme used in the CHIRRUP algorithm is shown in Figure~\ref{slot_patch}.

Concerning complexity, CHIRRUP makes $d\cdot 2^{p+r}$ calls to the chirp reconstruction algorithm, where $d$ is a small constant. Meanwhile the complexity of the chirp reconstruction algorithm was shown in Section~\ref{basic} to be $\mO[\tilde{n}K(K+\log_2^2 \tilde{n})]$, where $\tilde{n}:=2^m$. Combining these two observations, it follows that an analogous complexity result holds for CHIRRUP as for the chirp reconstruction algorithm, that is $\mO[nK(K+\log_2^2 n)]$.

\section{Numerical experiments}\label{experiments}

We wish to assess the ability of the CHIRRUP algorithm described in Sections~\ref{basic} and~\ref{enhancements} to decode a given number of messages. We choose to follow the framework proposed by Polyanskiy in~\cite{polyanskiy} in which the goal is to output a list of which messages were sent. Writing $\Lambda\subseteq\{1,\ldots,\mC\}$ for the set of transmitted messages, so that $|\Lambda|=K$, we seek an algorithm which outputs $\Gamma\subseteq\{1,\ldots,\mC\}$ such that $|\Gamma|=K$.\edit{\footnote{\edit{In our experiments we follow the framework in~\cite{polyanskiy} where it is assumed that the algorithm knows the number of messages $K$, which may not be practically realistic. It is worth noting, however, that our algorithm does not need to know the number of messages, and indeed its performance improves if we allow it to find more than $K$ messages.}}} Given $i\in\Lambda$, we write $E_i$ for the error event associated with message $i$, namely that $i\notin\Gamma$. In keeping with~\cite{polyanskiy}, we choose to assume an average-case model where the entries in $\Lambda$ are distributed uniformly at random on $\{1,\ldots,\mC\}$. We consider decoding to be successful if the \emph{per-user} probability of failure
$$\PPP:=\frac{1}{K}\sum_{i\in\Lambda}\PP(E_i)$$
is below some threshold $\epsilon$, where $\epsilon>0$ is a parameter to be chosen. An approach that has become popular~\cite{polyanskiy,ordentlich,krishna,krishna_CS} is then to study the trade-off between the transmission power $Q$ and the number of transmitted messages $K$, for a given failure probability threshold. One way to view transmission power is in terms of the average energy per bit required to transmit each message, denoted by $E_b/N_0$, and defined to be
$$\frac{E_b}{N_0}:=\frac{nQ}{2B}.$$

We plot minimum required $E_b/N_0$ against $K$ for our enhanced chirp reconstruction to achieve a success probability of above $0.95$ ($\epsilon=0.05$). We fix the code length\footnote{This corresponds to a real codelength of $2^{15}=32,768$.} to be $n=2^{14}=16,384$. Various slotting and patching configurations as described in Section~\ref{enhancements} are possible for the CHIRRUP algorithm: we vary the number of slots $2^p$ and number of patches $2^r$ for different positives integers $p$ and $r$. We restrict to parameter choices for which the number of bits exceeds $40$, for which the number of permissible messages exceeds $80$, and for which the algorithm running time is not prohibitive. We find that, generally speaking, a choice of $p\in\{5,6,7\}$ is required to meet these conditions: larger $p$ leads to too short or too few messages, while smaller $p$ leads to too few messages or prohibitive run time.\footnote{\edit{Since the number of messages decodable by the chirp reconstruction algorithm grows sublinearly with codebook length, when $p$ is small the number of messages decreases. On the other hand, if $p$ is taken too large and the codebook length is very small, the chirp reconstruction algorithm ceases to perform well.}} We also need to choose the number of parity check bits when the number of patches exceeds $1$: we find by experimentation that $l=\begin{bmatrix}0&15\end{bmatrix}$ and $l=\begin{bmatrix}0&10&10&15\end{bmatrix}$ are sensible choices for $r=1$ and $r=2$ respectively. These choices are made with a view to minimizing the number of bits devoted to parity checking whilst keeping the probability of information loss in the patching process low. We find that only a modest number of patches is beneficial (up to $4$), which deviates somewhat from the implementation in~\cite{krishna_CS} in which $11$ patches were used. It should be noted that the much more computationally demanding nonnegative least squares is used as the reconstruction algorithm in~\cite{krishna_CS}, which necessitates the splitting up of messages into more patches to make the reconstructions computationally tractable.

\begin{figure}
\centering
\begin{subfigure}[b]{0.45\textwidth}{\includegraphics[width=\textwidth]{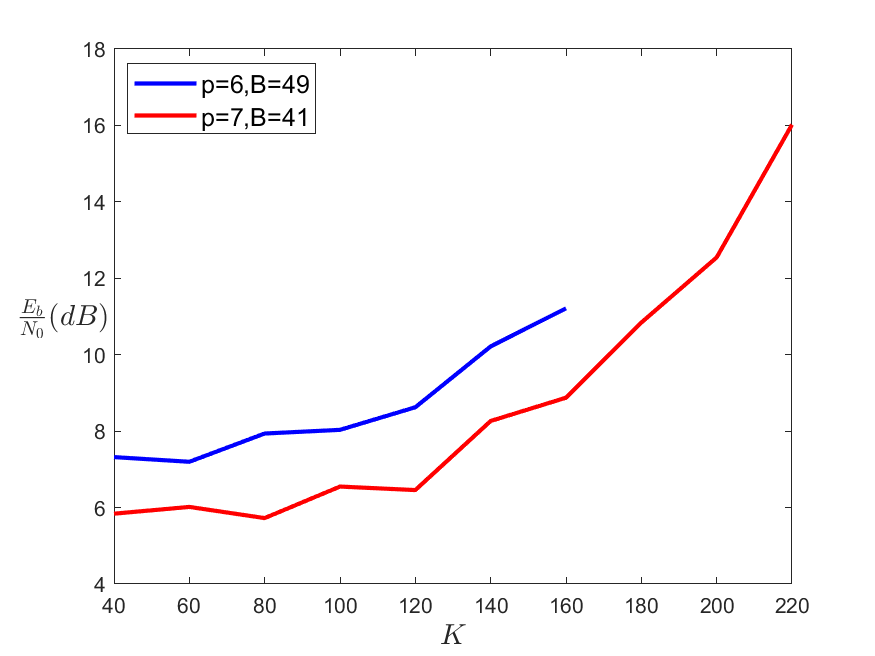}\caption{$E_b/N_0$ for 1 patch.}\label{EbN0_p1}}\end{subfigure}
\begin{subfigure}[b]{0.45\textwidth}{\includegraphics[width=\textwidth]{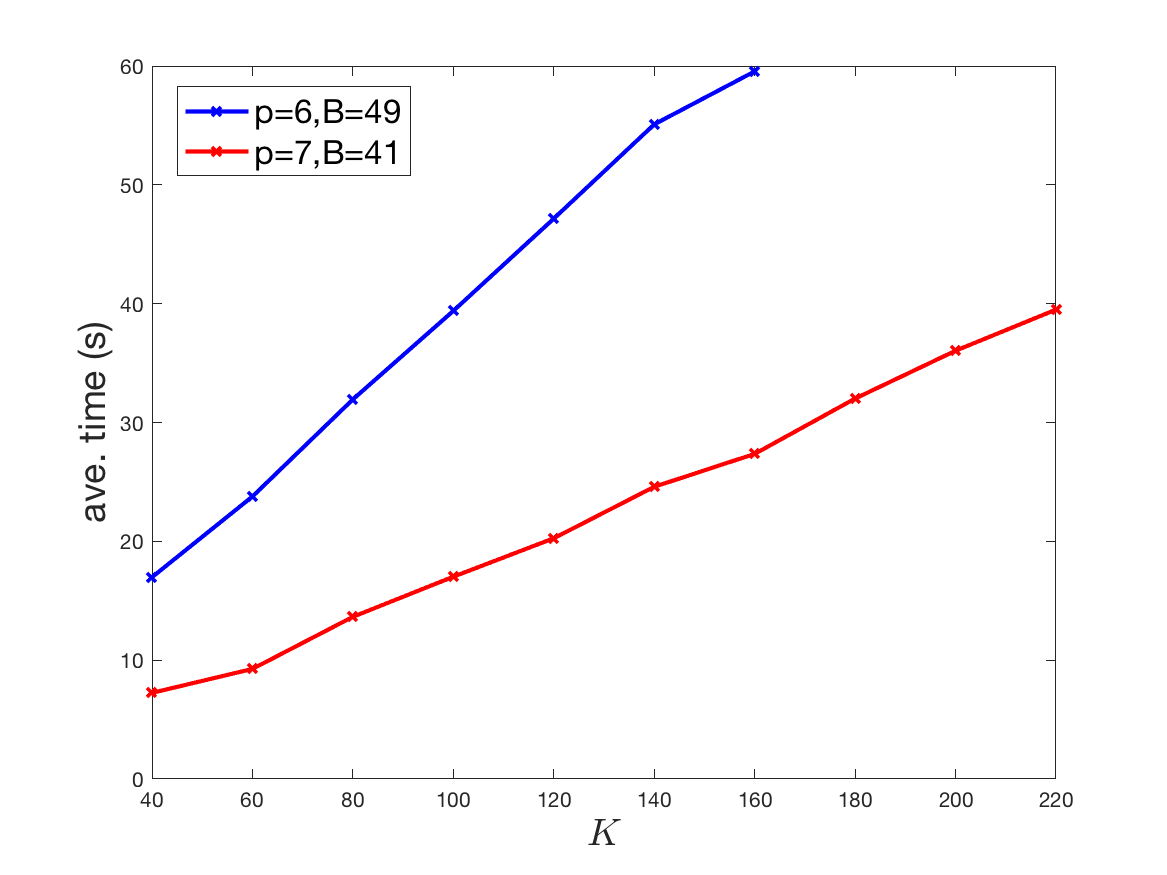}\caption{Running time for 1 patch.}\label{time_p1}}\end{subfigure}
\begin{subfigure}[b]{0.45\textwidth}{\includegraphics[width=\textwidth]{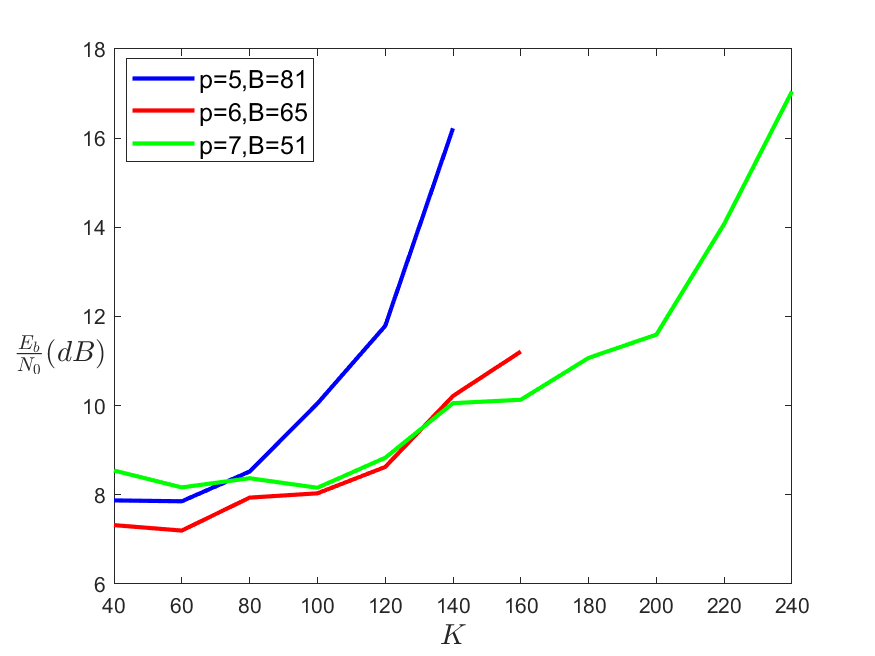}\caption{$E_b/N_0$ for 2 patches.}\label{EbN0_p2}}\end{subfigure}
\begin{subfigure}[b]{0.45\textwidth}{\includegraphics[width=\textwidth]{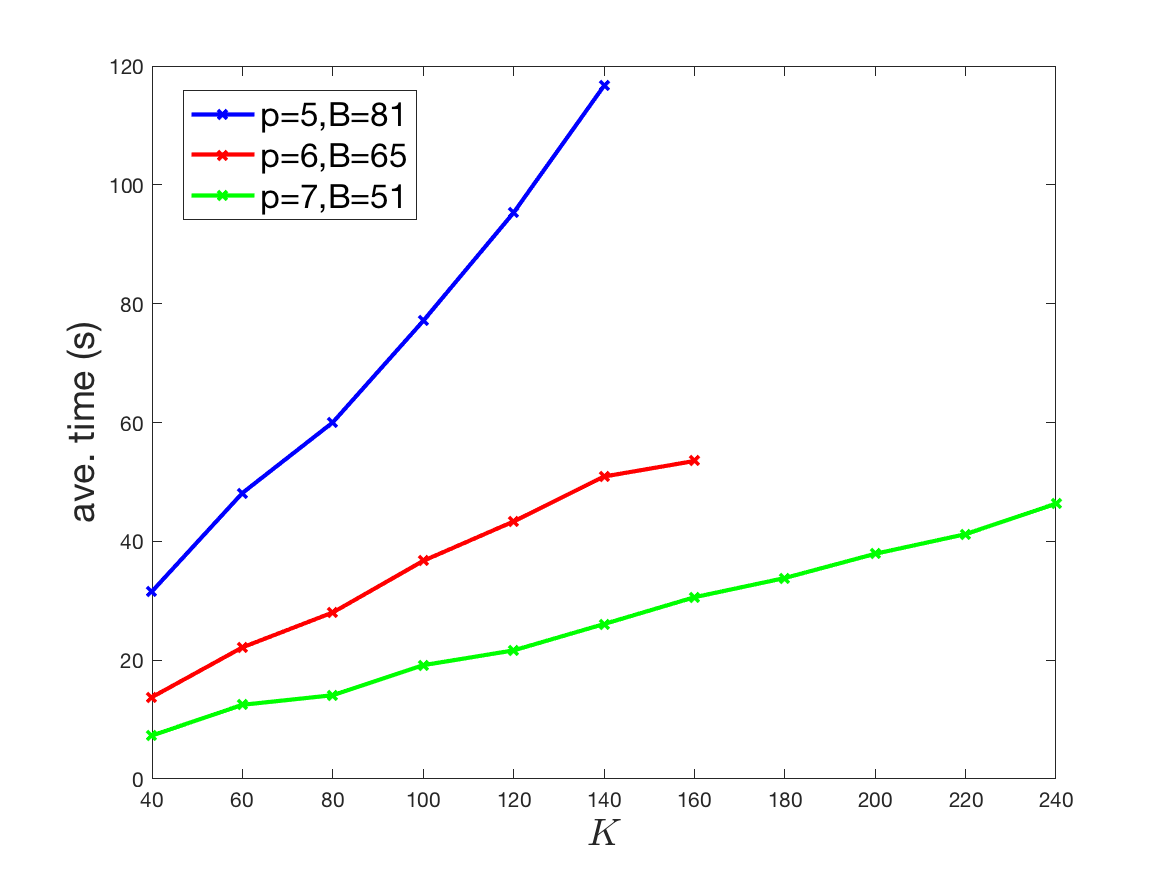}\caption{Running time for 2 patches.}\label{time_p2}}\end{subfigure}
\begin{subfigure}[b]{0.45\textwidth}{\includegraphics[width=\textwidth]{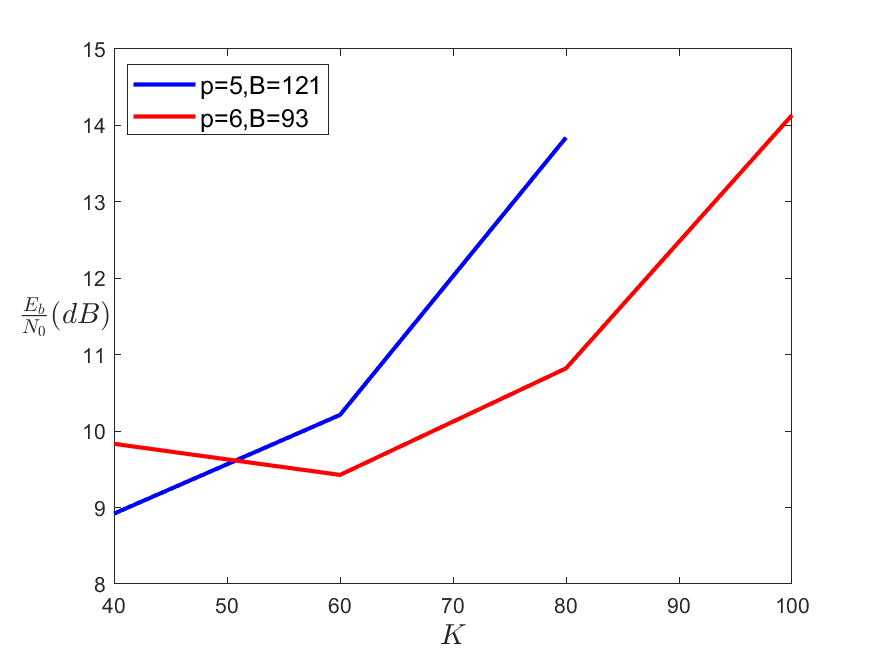}\caption{$E_b/N_0$ for 4 patches.}\label{EbN0_p4}}\end{subfigure}
\begin{subfigure}[b]{0.45\textwidth}{\includegraphics[width=\textwidth]{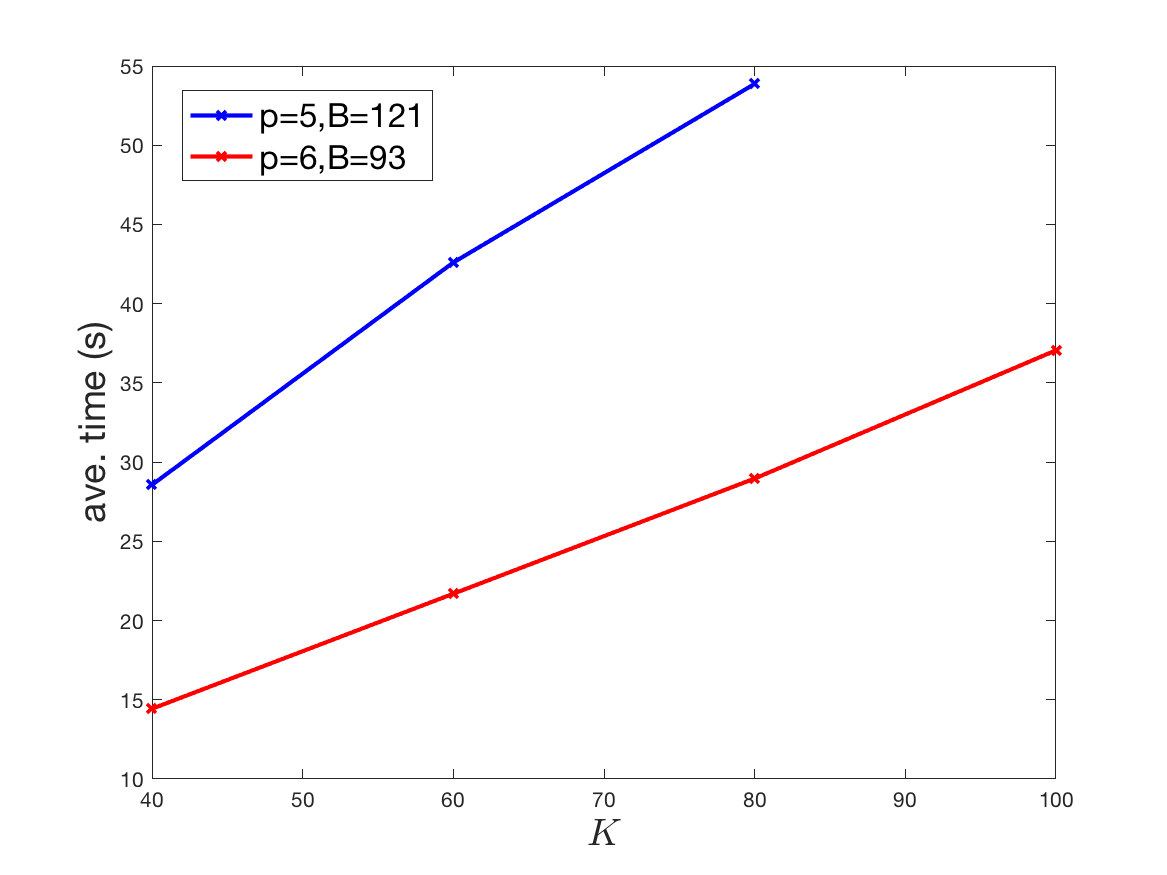}\caption{Running time for 4 patches.}\label{time_p4}}\end{subfigure}
\caption{Energy per bit (subfigures (a), (c) and (e)) and average computation time (subfigures (b), (d) and (f)) for CHIRRUP with complex chirps.\label{EbN0}}
\end{figure}

Plots of $E_b/N_0$ (displayed on a decibel scale) against $K$ are shown for $1$, $2$ and $4$ patches in Figures~\ref{EbN0_p1},~\ref{EbN0_p2} and~\ref{EbN0_p4} respectively. We observe that the slotting parameter $p$ allows for a trade-off between both number of transmitted messages and energy per bit ($E_b/N_0$) on the one hand and message length ($B$) on the other hand. As the number of slots is increased, more messages can be decoded and $E_b/N_0$ generally decreases for a given $K$. This gain is achieved at the cost of reducing the codeword size within each slot, which decreases the message length. 

We also observe that increasing the number of patches, while keeping the number of slots fixed, both increases the message length and decreases the number of messages that can be decoded, thus representing a second trade-off. As a rough guide, $1$ patch appears to be preferable for modest bit lengths of under $50$ bits, $2$ patches are preferable for between $50$ and $80$ bits and $4$ patches are preferable for upwards of $80$ bits.

We observe in both Figures~\ref{EbN0_p2} and ~\ref{EbN0_p4} that the curves for $p=5$ and $p=6$ cross each other, which might appear surprising. The authors believe that the explanation for this behaviour lies in suboptimal tuning of a certain parameter, namely how `close' in value to $1$ a coefficient must be for its corresponding chirp component to be retained (see Section~\ref{message}). One of the advantages of CHIRRUP is that no tuning of parameters is required: the same parameter choices were used throughout all the tests. However, our observation is that the threshold selected is somewhat suboptimal for the case of a small number of messages (e.g. $K=40$), and this accounts for the unusual crossing behaviour.

Average computation time is displayed for $1$, $2$ and $4$ patches in Figures~\ref{time_p1},~\ref{time_p2} and~\ref{time_p4} respectively. For a given number of messages $K$, the energy per bit $E_b/N_0$ is set to the smallest which permits a success probability of $0.95$ as determined by our experiments. The computations were carried out on a 2.2 GHz MacBook Air. We observe that the CHIRRUP algorithm can be run in a few seconds with desktop computing power. We observe a tradeoff between message length and running time: longer messages require larger binary chirp codebooks, which makes the algorithm more computationally burdensome.

\begin{figure}
\centering
\begin{subfigure}[b]{0.45\textwidth}{\includegraphics[width=\textwidth]{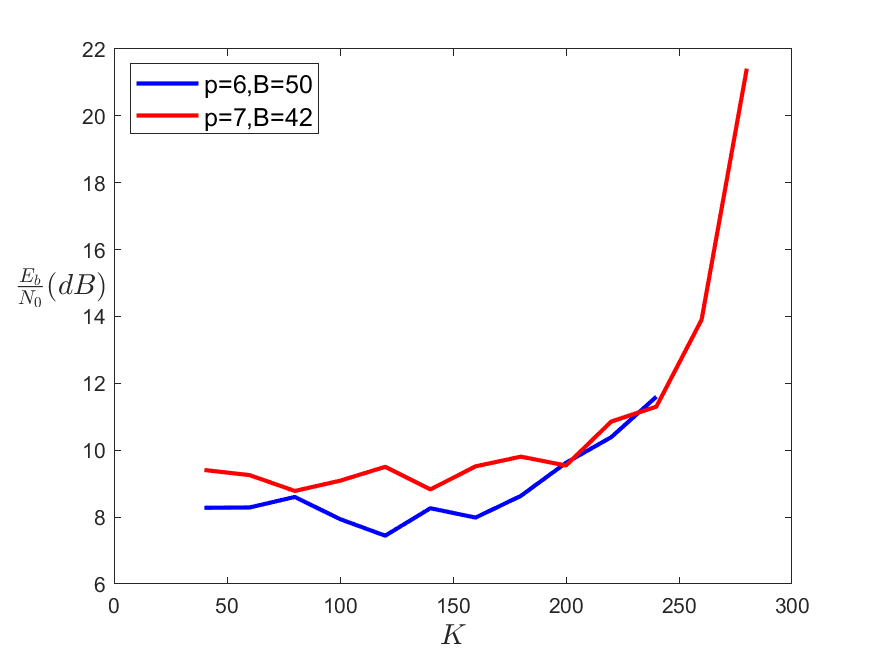}\caption{$E_b/N_0$ for 1 patch.}\label{EbN0_p1R}}\end{subfigure}
\begin{subfigure}[b]{0.45\textwidth}{\includegraphics[width=\textwidth]{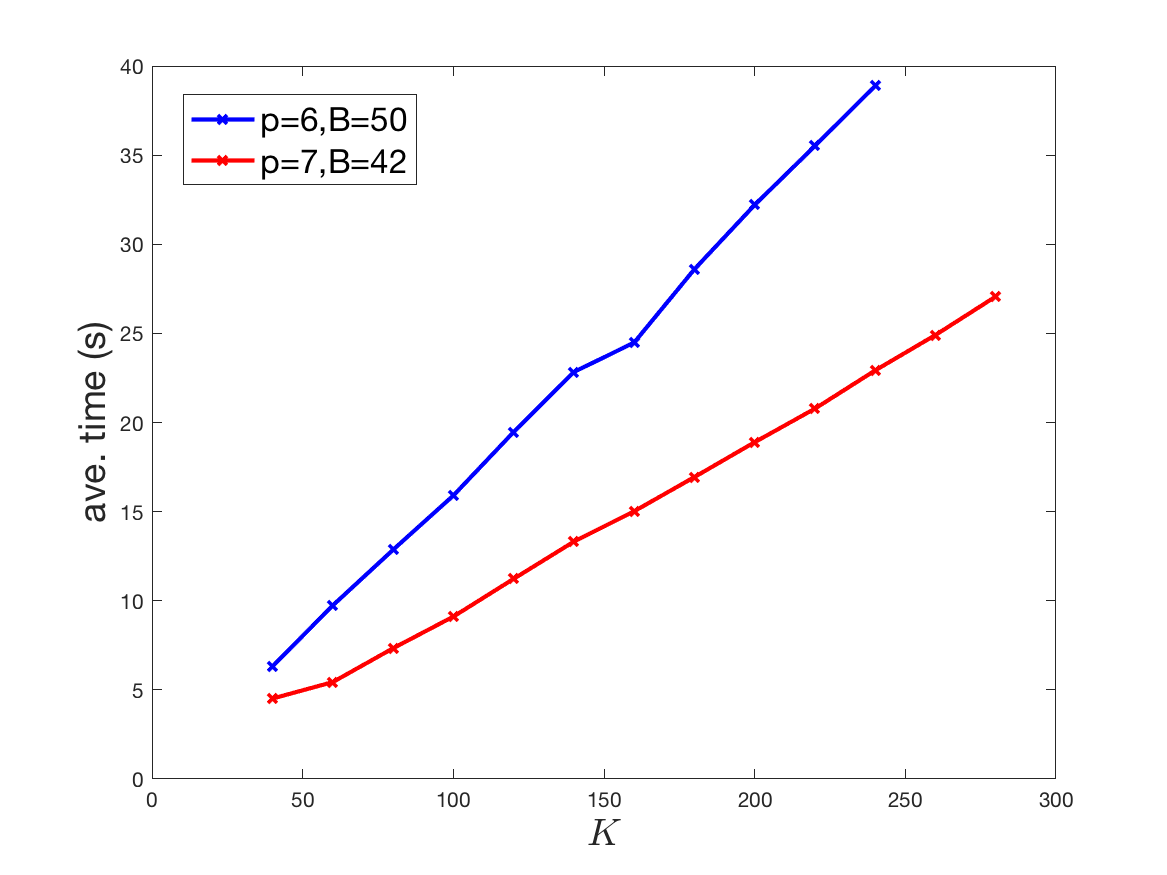}\caption{Running time for 1 patch.}\label{time_p1R}}\end{subfigure}
\begin{subfigure}[b]{0.45\textwidth}{\includegraphics[width=\textwidth]{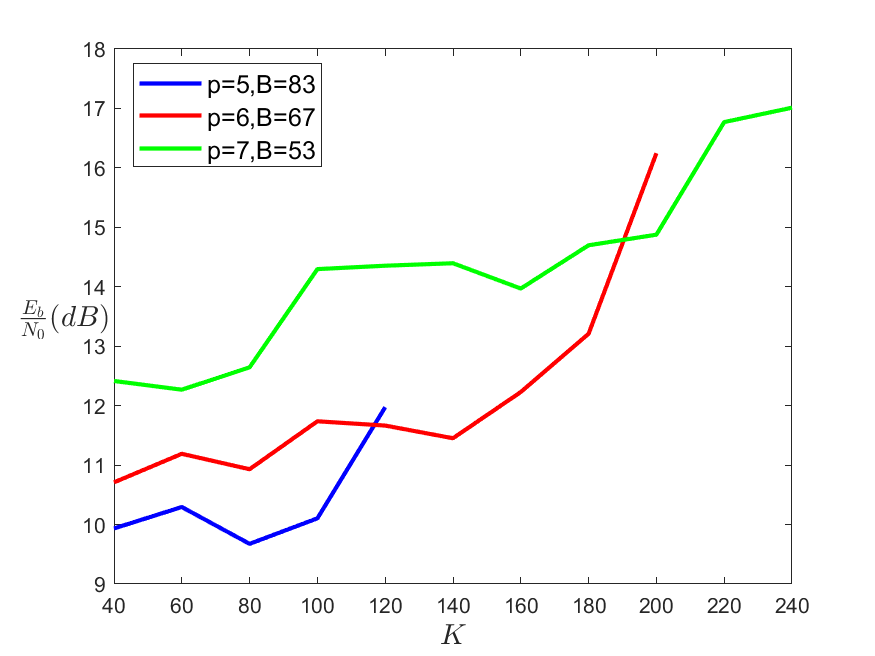}\caption{$E_b/N_0$ for 2 patches.}\label{EbN0_p2R}}\end{subfigure}
\begin{subfigure}[b]{0.45\textwidth}{\includegraphics[width=\textwidth]{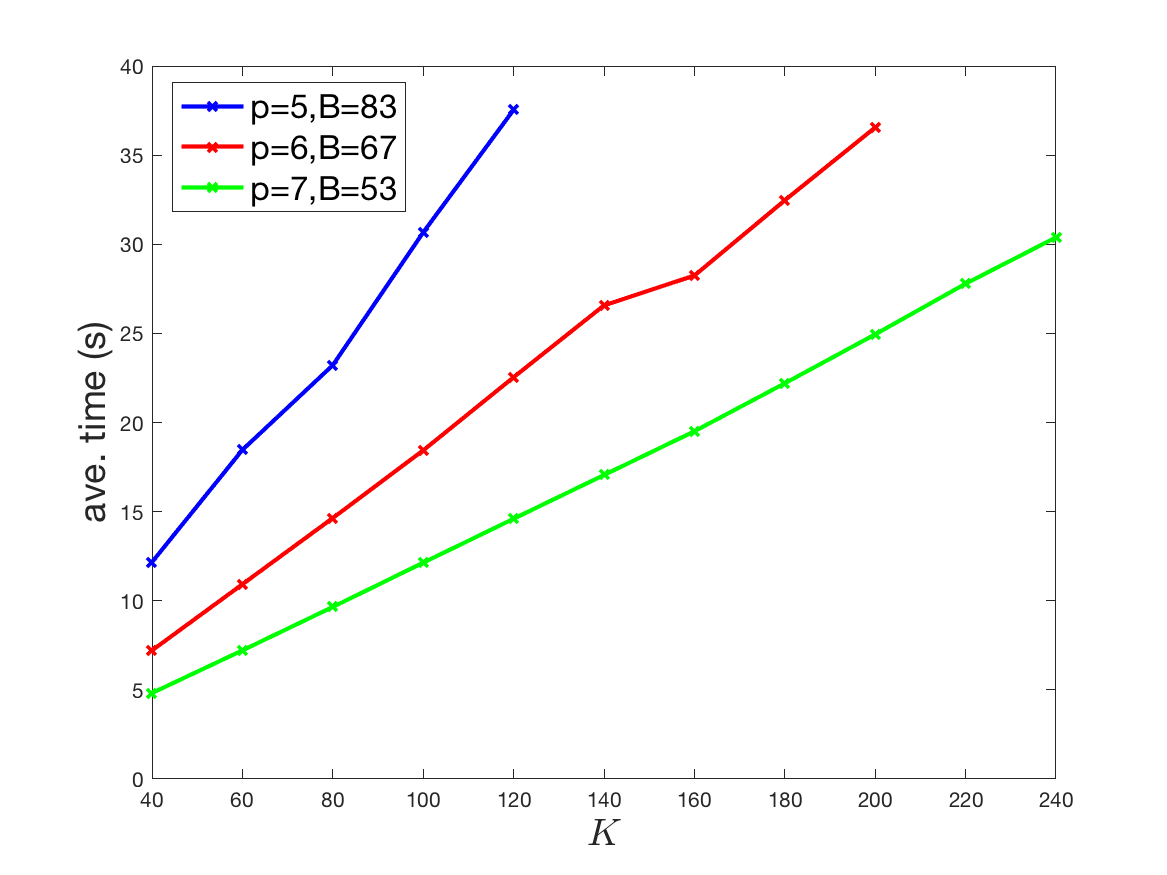}\caption{Running time for 2 patches.}\label{time_p2R}}\end{subfigure}
\begin{subfigure}[b]{0.45\textwidth}{\includegraphics[width=\textwidth]{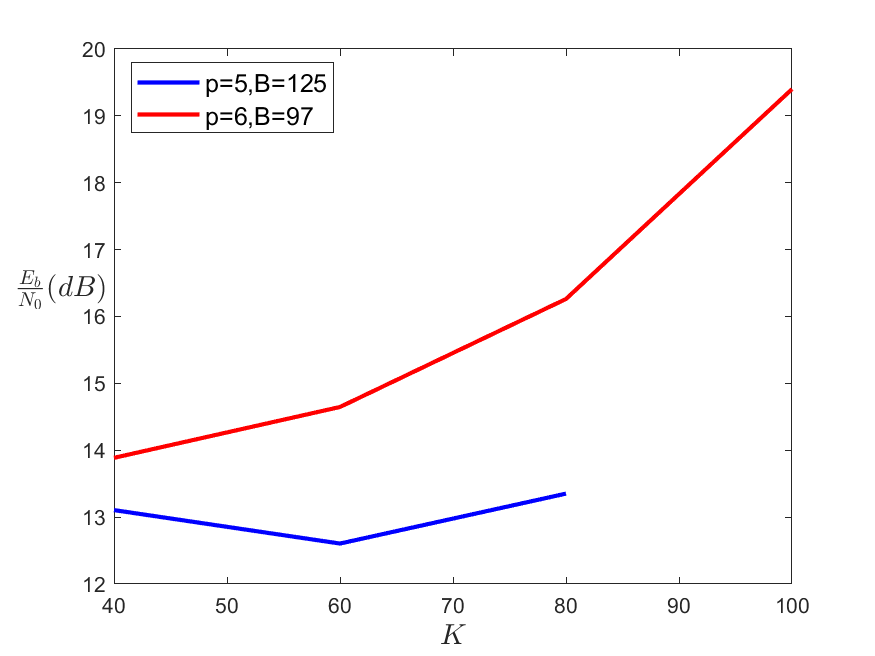}\caption{$E_b/N_0$ for 4 patches.}\label{EbN0_p4R}}\end{subfigure}
\begin{subfigure}[b]{0.45\textwidth}{\includegraphics[width=\textwidth]{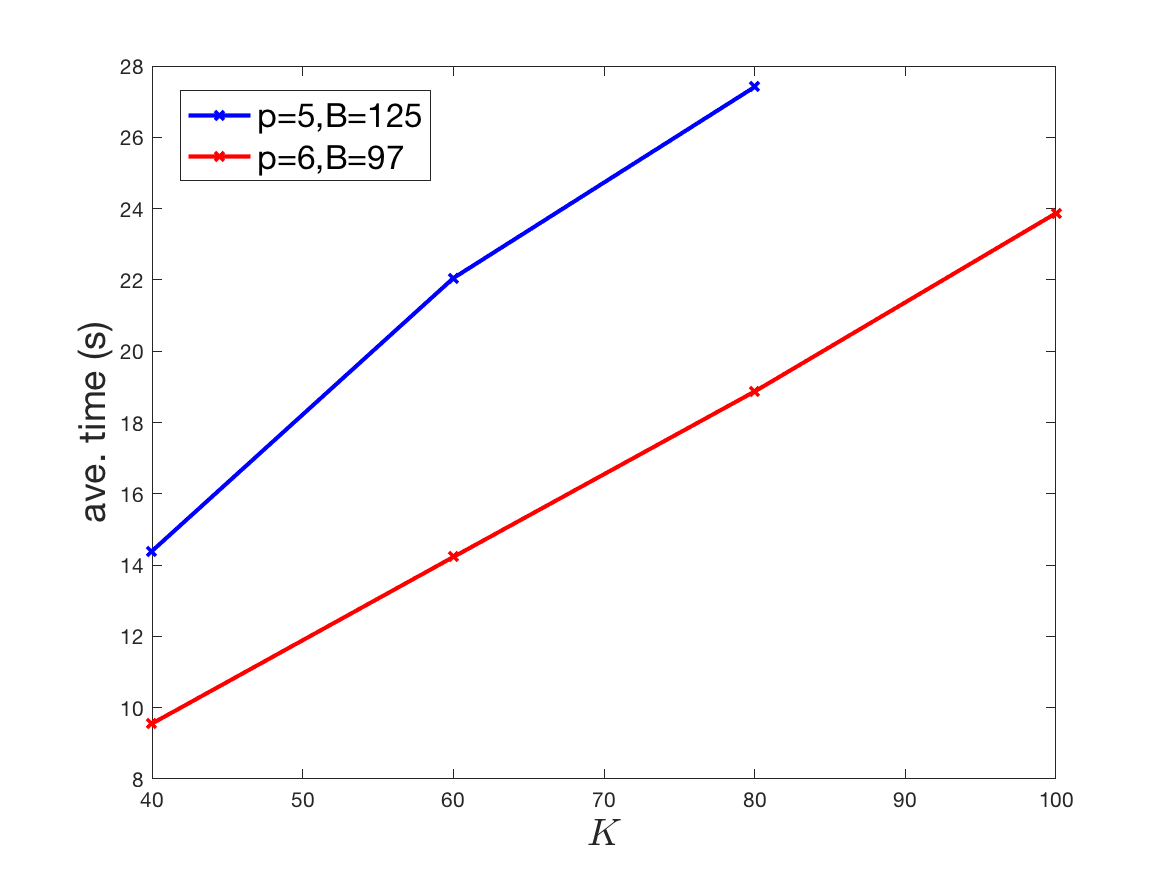}\caption{Running time for 4 patches.}\label{time_p4R}}\end{subfigure}
\caption{Energy per bit (subfigures (a), (c) and (e)) and average computation time (subfigures (b), (d) and (f)) for CHIRRUP with real chirps.\label{EbN0_real}}
\end{figure}

\paragraph{Real codebooks.} We noted that a codelength of $n=2^{14}$ corresponds to a real codelength of $2^{15}$. Rather than using complex binary chirps, an alternative approach is to restrict to real binary chirps from the outset. In this case, a fair comparison is to take $n=2^{15}$. We performed analogous numerical experiments for real codebooks as for complex codebooks. Plots of energy per bit against number of messages are displayed for $1$, $2$ and $4$ patches in Figures~\ref{EbN0_p1R},~\ref{EbN0_p2R} and~\ref{EbN0_p4R} respectively, and plots of running time against number of messages for $1$, $2$ and $4$ patches are displayed in Figures~\ref{time_p1},~\ref{time_p2} and~\ref{time_p4} respectively. 

We observe a somewhat close correspondence between the results for complex and real binary chirps, which reflects the fact that the corresponding codebooks are geometrically equivalent. However, for some parameter choices real chirps allow for a better tradeoff between number of bits and number of messages, especially for $r=0$, and to some extent for $r=1$. We believe that this improvement is not due to a fundamental difference between complex and real binary chirp codebooks, but rather to a different condition for deciding whether the coefficient corresponding to a retrieved component is close to $1$. The results therefore emphasize how sensitive the algorithm is to the choice of this condition, and that it certainly pays to make use of the prior knowledge that all signal coefficients are equal to $1$. A thorough investigation of how to optimize the condition is unfortunately a computationally intimidating task, which prevented further exploration of this behaviour.

\section{Theoretical comparison: One Step Thresholding}\label{theory}

Among the plethora of algorithms for compressed sensing reconstruction, the CHIRRUP algorithm is one of the few with the distinction of having running time sublinear in the signal dimension. It is nonetheless interesting to know how the performance of  CHIRRUP would compare to other algorithms which scale poorly in this context. In this regard, we choose the very simple One Step Thresholding (OST) algorithm~\cite{why_gabor}, which is also known as \emph{treat interference as noise} (TIN)~\cite{polyanskiy}. The algorithm\footnote{The OST algorithm described in~\cite{why_gabor} considers the absolute values of the $\{g_i\}$. We remove the absolute value since we have prior knowledge that all signal coefficients are positive.} is stated succinctly in Algorithm~\ref{OST}.
\begin{algorithm}\caption{One Step Thresholding~\cite{why_gabor}\label{OST}}
\textbf{Inputs}: $y\in\CC^n$, $X\in\CC^{n\times\mC}$, $K$.
\begin{enumerate}
\item $g=X^{\ast} y$.
\item $\Gamma:=\{i\;\textrm{corresponding to the}\;K\;\textrm{largest}\;g_i\}.$
\end{enumerate}
\textbf{Outputs}: $\Gamma$.
\end{algorithm}
Assuming that the entries of the codebook matrix $X$ are i.i.d. zero-mean Gaussian, it can be shown that, as problem size grows, the entries of $g$ themselves converge in distribution to a Gaussian distribution, with nonzero mean for the entries corresponding to transmitted messages and zero mean for the other entries. The precise behaviour of OST can thus be determined in the asymptotic limit. Writing
$$\delta=\lim_{K\rightarrow\infty}\frac{K}{\mC}\;\;\;\textrm{and}\;\;\;\rho=\lim_{K\rightarrow\infty}\frac{K+1/Q}{n},$$
we find that, for a given failure probability $\epsilon$ (see Section~\ref{experiments}), there exists a phase transition threshold in terms of $\delta$ and $\rho$. Note the striking interpretation: for a given problem size, OST is subject to a fundamental limit on the sum of the number of messages and the inverse of the power. This quantity $K+1/Q$ can be viewed as an `effective number of messages', which is a combination of the true number of messages and the equivocation arising from the noise. The asymptotic behaviour of OST for random coding has been previously derived in~\cite{verdu} using an information-theoretic approach and quantified in~\cite{polyanskiy,ordentlich}. In Appendix~\ref{OST_asympt}, we provide an alternative derivation which highlights the ``inverse of signal power = effective sparsity'' interpretation.

Moreover, it has been observed that the average-case behaviour of many compressed sensing algorithms remains indistinguishable when Gaussian random codebooks are replaced with many other deterministic codebooks, including Delsarte-Goethals frames which are built from binary chirps~\cite{deterministic_phase}. These asymptotic results for Gaussian matrices would therefore be expected to also accurately predict the performance of OST with binary chirp codebooks.

\begin{figure}[t!]
\centering
\includegraphics[width=0.8\textwidth]{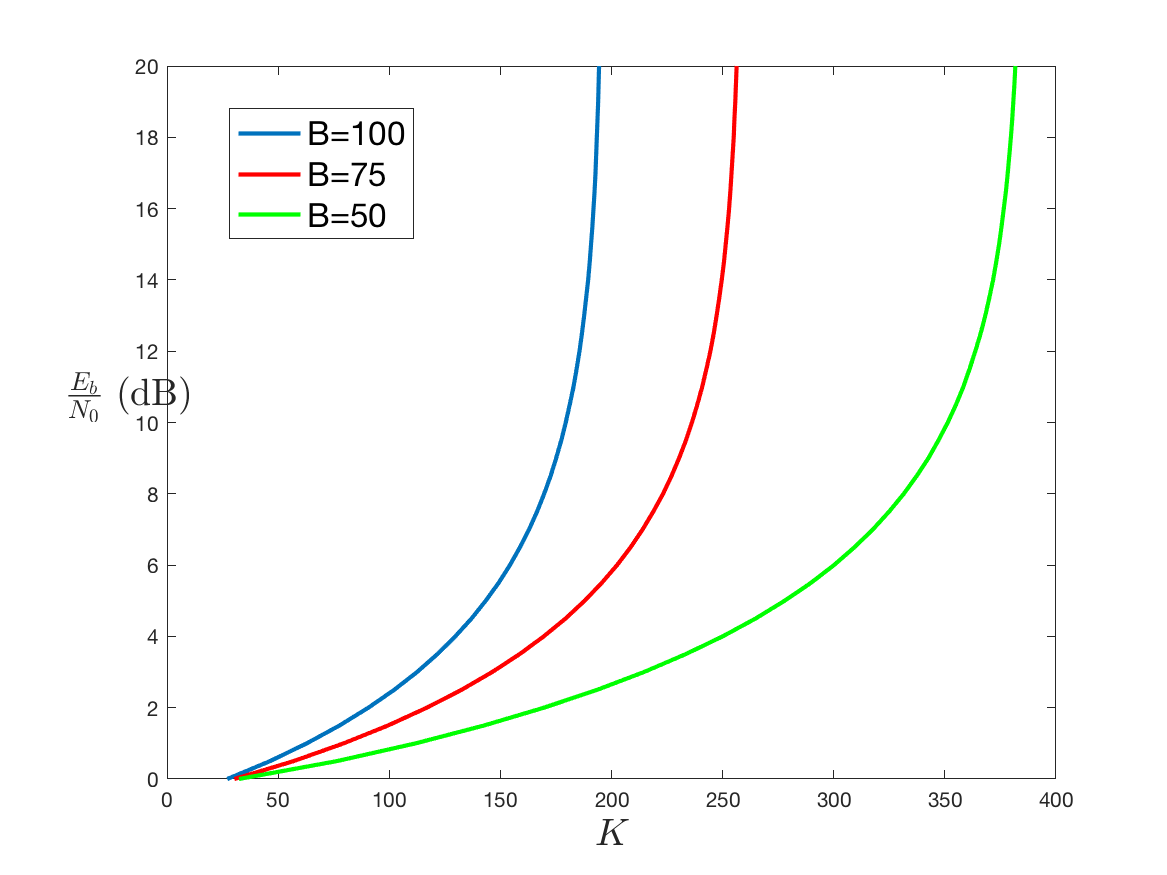}
\caption{Expected performance of OST for different numbers of messages.}
\label{OST_compare}
\end{figure}

We choose the same problem dimensions as in our numerical tests: $n=2^{15}=32,768$ and error probability $\epsilon=0.05$. Using the Gaussian asymptotic approximation, we obtain predictions for the performance of OST for varying numbers of bits ($B=50$, $B=75$ and $B=100$) in Figure~\ref{OST_compare}. We observe that the observed performance of CHIRRUP is comparable to OST, with number of transmitted messages easily within a factor of $2$. This is despite the fact that CHIRRUP is sublinear in the signal dimension, in comparison with OST which is superlinear, in fact $\mO(\mC\log\mC)$. It should be emphasized that the plots in Figure~\ref{OST_compare} are hypothetical: without enormous computational resources the OST algorithm is completely intractable for the problems addressed in this paper.

\section{Conclusions and future work}\label{conclusions}

\subsection{Conclusions}

We have presented CHIRRUP, an algorithm for unsourced multiple access based upon binary chirp coding and the chirp reconstruction algorithm~\cite{chirp_reconstruction}. We have presented the results of numerical simulations, using problem sizes which are realistic for applications, detailing how energy per bit ($E_b/N_0$) and computing time depends upon number of active users. We believe that this work is the first to engage quantitatively with the performance versus computational efficiency tradeoff for this problem. By contrast, prior contributions either stop short of proposing an actual algorithm (making use of optimal coding bounds for finite blocklength codes)~\cite{ordentlich,krishna} or propose an algorithm without quantifying computational efficiency~\cite{krishna_CS}. We suggest the use of CHIRRUP as a benchmark against which other practical algorithms can be compared numerically. We have made available \texttt{MATLAB} code for CHIRRUP at \texttt{https://github.com/ajthompson42/CHIRRUP}.

Viewed another way, CHIRRUP can also be viewed as a combined deterministic measurement scheme and reconstruction algorithm for compressed sensing at massive scale. We have pointed out that CHIRRUP is sublinear in the signal dimension in contrast to almost all other compressed sensing algorithms. For example, not one of the algorithms surveyed in a much-cited review of algorithms for sparse linear inverse problems~\cite{linear_inverse} has sublinear complexity. In this respect, CHIRRUP occupies a unique place among compressed sensing algorithms in its ability to scale to problems at massive scale.

\subsection{Future work}

\paragraph{Extension to witness-averaging.} The \emph{witness-averaging} algorithm, presented in~\cite{witness}, is a modified version of the chirp reconstruction algorithm (Algorithm~\ref{chirp_alg}). In this approach, a \emph{Reed-Muller Sieve} codebook is used which consists of binary chirps given by all binary symmetric matrices $P$ but without the phase shifts $b$. Most significantly, instead of summing two shift-and-multiply computations as in the present paper, all possible shifts are computed and summed. This change has the effect of increasing the complexity of the algorithm so that it is no longer sublinear -- in fact $\mO(\mC\log\mC)$ -- while at the same time drastically improving its decoding performance so that $\mO(n)$ components can be recovered. It would be interesting to investigate how the introduction of some measure of witness-averaging affects the performance of CHIRRUP on the number of messages versus computational time tradeoff. It is worth noting that such a change would move CHIRRUP closer in spirit to the approach in~\cite{krishna_CS}, in which a more computationally demanding but better performing compressed sensing algorithm is employed over a larger number of message patches.

\paragraph{Application to neighbor discovery.} In conventional networks, the overhead of neighbor discovery can be amortized over long data transmissions. However, when there is a massive number of devices, and when transmissions are short and bursty, it has been shown that the overhead of neighbor discovery may considerably reduce data throughput~\cite{gaussian}.

Neighbor discovery can be formulated as multiuser detection~\cite{neighbor} and the coupled compressive sensing approach of~\cite{krishna_CS} was recently extended to the problem of asynchronous neighbor discovery in~\cite{krishna_neighbor}. Since the number of active devices is typically orders of magnitude smaller than the total number of devices, it is natural to apply ideas from compressed sensing. Synchronous transmission is considered in~\cite{sparse_multiuser}, and asynchronous transmission in~\cite{asynchronous_CDRA}, where the LASSO algorithm is used to detect active devices. Note that the computational complexity of LASSO is polynomial in the total number of devices. Codewords in~\cite{asynchronous_CDRA} are obtained by exponentiating codewords in a 2nd order Reed-Muller code. It is natural to explore replacing LASSO by chirp reconstruction, since this would significantly improve efficiency.

The term \emph{random access} is commonly used to describe a setup where a random subset of users in the network communicate with a base station (BS) in an uncoordinated fashion. The on-off random access channel represents an abstraction where users signal activity by transmitting their identity~\cite{on_off}. The primary objective of the BS is to recover the set of active users. Multiuser detection via One Step Thresholding (OST) is considered in~\cite{on_off,why_gabor}.

When a wireless device is equipped with half-duplex hardware, it can either transmit or receive signals in a given time slot, but it cannot do both simultaneously. A method of achieving full-duplex communication using half-duplex radius is proposed in~\cite{RODD}. It is called random on-off-division duplex (RODD) becuase each user is assigned a unique on-off sequence, and during the off slots each user listens to the signals transmitted by the other users.

A key question in this context is the design of a suitable codebook of on-off sequences. It must be amenable to effective and efficient decoding using compressed sensing algorithms and it must be sparse. Luo and Guo~\cite{group_testing,rayleigh} proposed using random Bernoulli codebooks along with a group testing reconstruction algorithm. Subsequently Zhang and Guo proposed codebooks based on second order Reed-Muller codes along with a modified chirp reconstruction algorithm~\cite{neighbor_CS}.

The present authors proposed sparse Kerdock matrices as codebooks for the neighbor discovery problem~\cite{kerdock_neighbour}. These matrices share the same row space as certain Delsarte-Goethals frames, while at the same time being extremely sparse. Numerical experiments with One Step Thesholding (OST) and Normalized Iterative Hard Thresholding (NIHT) demonstrate that a higher proprtion of neighbors are successfully identified using sparse Kerdock matrices than with codebooks obtained by randomly erasing entries in a Reed-Muller codebook (as proposed in~\cite{neighbor_CS}). It is natural to consider replacing OST and NIHT by chirp reconstruction.

\appendix

\section{Proof of Proposition~\ref{shift_multiply}}\label{shift_multiply_proof}

Recalling~(\ref{chirp_def}), we have
\begin{eqnarray}
\overline{\left\{\phi_{P_k,b_k}\right\}_a}\left\{\phi_{P_l,b_l}\right\}_{a+e}&=&Qi^{-2b_k^T a-a^TP_k a}i^{2b_l^T(a+e)+(a+e)^TP_l(a+e)}\nonumber\\
&=&Qi^{2b_l^Te+e^TP_l e+2(b_l-b_k+P_l e)^T a+a^T(P_l-P_k)a}\nonumber\\
&=&\left\{\phi_{P_l,b_l}\right\}_e\left\{\phi_{P_l-P_k,b_l-b_k+P_l e}\right\}_a.\label{cross_terms}
\end{eqnarray}
When $k\neq l$, we have $P_k\neq P_l$, and (\ref{cross_terms}) consists of a constant term (a chirp in $e$) multiplied by a chirp in $a$. However, when $k=l$, we have $P_k=P_l$ and (\ref{cross_terms}) reduces to
\begin{equation}\label{matched_terms}
\overline{\left\{\phi_{P_k,b_k}\right\}_a}\left\{\phi_{P_k,b_k}\right\}_{a+e}=\sqrt{Q}\left\{\phi_{P_k,b_k}\right\}_e(-1)^{e^T P_k a}.
\end{equation}
Substituting (\ref{cross_terms}) and (\ref{matched_terms}) into (\ref{measurements}), we obtain
\begin{eqnarray*}
f_a=\overline{y_a}y_{a+e}&=&\overline{\left[\sum_{k=1}^K \phi_{P_k,b_k}+z\right]_a}\left[\sum_{l=1}^K \phi_{P_l,b_l}+z\right]_{a+e}\\
&=&\sum_{k=1}^K\overline{\left\{\phi_{P_k,b_k}\right\}_a}\left\{\phi_{P_k,b_k}\right\}_{a+e}+\sum_{k\neq l}\overline{\left\{\phi_{P_k,b_k}\right\}_a}\left\{\phi_{P_l,b_l}\right\}_{a+e}\\
&&\;\;\;\;+\overline{z}_a \left[\sum_{k=1}^K \phi_{P_k,b_k}\right]_{a+e}+z_{a+e}\overline{\left[\sum_{k=1}^K \phi_{P_k,b_k}\right]_a}+\overline{z}_a z_{a+e}\\
&=&\sqrt{Q}\sum_{k=1}^K\left\{\phi_{P_k,b_k}\right\}_e(-1)^{e^T P_k a}+\sum_{k\neq l}\left\{\phi_{P_l,b_l}\right\}_e\left\{\phi_{P_l-P_k,b_l-b_k+P_l e}\right\}_a\\
&&\;\;\;\;+\overline{z}_a\left[\sum_{k=1}^K \phi_{P_k,b_k}\right]_{a+e}+z_{a+e}\overline{\left[\sum_{k=1}^K \phi_{P_k,b_k}\right]_a}+\overline{z}_a z_{a+e},
\end{eqnarray*}
as required.

\section{Gaussian approximation for OST}\label{OST_asympt}

As in Section~\ref{intro}, we consider measurements of the form
$$y=Xs+z$$
where $s=\begin{bmatrix}s_1&\dots&s_{\mC}\end{bmatrix}^T\in\{0,1\}^{\mC}$ is the activity pattern with Hamming weight $K$ and $z\sim\mathcal{N}(0,I_n)$ is additive white Gaussian noise (AWGN). We now assume that the entries of $X$ are distributed i.i.d $N(0,Q)$. We derive the asymptotic distribution of the entries of $g$, as defined in Algorithm~\ref{OST}, under the given model for $X$ and $z$. More precisely, we consider a sequence of problems with parameters $(K_r,n_r,\mC_r)$ for $r=1,2,\ldots$ such that
$$\frac{K_r}{\mC_r}\rightarrow\delta,\;\;\;\;\frac{K_r+1/Q}{n_r}\rightarrow\rho.$$
Writing $X_i$ for the $i$th column of $X$, we note that $\|X_i\|_2^2/n\sim\frac{Q}{n}\chi_2^n$ which converges in distribution exponentially fast to $Q$ by the law of large numbers as $n\rightarrow\infty$, and so $Q$ continues to represent the transmitted power per message (see Section~\ref{intro}). Writing $G_i$ for the random variable corresponding to the entries $g_i$, we have a central limit result for each of the $G_i$.

\begin{theorem}\label{central_limit}
$$\frac{G_i}{nQ}\xrightarrow{p}\left\{\begin{array}{ll}
N(1,\rho)&i\in\Lambda\\
&\\
N(0,\rho)&i\in\Lambda^C.\end{array}\right.$$
\end{theorem}
\textbf{Proof:} Fix $(K,n,Q)$ and suppose $p\in\Lambda^C$ and consider the random variable
$$Y_i^n:=\frac{1}{Q\sqrt{n}}X_{ip}\left(\sum_{j\in\Lambda}X_{ij}+z_i\right).$$
Note that the $\{Y^n_i\}$ are independent for $i=1,\ldots,n$, since they involve entries from different rows of $X$. Furthermore, we have $\EE Y_i=0$ and $\EE Y_i^2=\frac{K+1/Q}{n}<\infty$. Then by the Berry-Esseen theorem the distribution function, $F_n(x)$, of
$$\frac{\displaystyle\sum_{i=1}^n Y_i}{\sqrt{n\,\EE Y_i^2}}=\frac{\displaystyle\sum_{i=1}^n X_{ip}\left(\sum_{j\in\Lambda}X_{ij}+z_i\right)}{nQ\left(\frac{K+1/Q}{n}\right)^{1/2}}$$
satisfies
\begin{equation}\label{berry_esseen}
\sup_x|F_n(x)-\Phi(x)|\le\frac{\EE |Y_i|^3}{\sqrt{n}\left(\frac{K+1/Q}{n}\right)^{3/2}},
\end{equation}
where $\Phi(x)$ is the distribution function of the standard Normal distribution. It remains to calculate $\EE|Y_i|^3$. In this regard, we note that, if $Z\sim N(0,\sigma^2)$, we have
$$\EE|Z|^3=\frac{2\sqrt{2}\sigma^3}{\sqrt{\pi}}.$$
Now, for all $i=1,\ldots,n$,
$$X_{ip}\sim N\left(0,Q\right),\;\;\;\;\textrm{and}
\;\;\;\;\sum_{j\in\Lambda}X_{ij}+z_i\sim N\left(0,KQ+1\right),$$
and they are mutually independent. We therefore have
$$\begin{array}{rcl}
\EE|Y_i^n|^3&=&Q^{-3}n^{-3/2}\EE|X_{ip}|^3\,\EE\left(\left|\sum_{j\in\Lambda}X_{ij}+z_i\right|^3\right)\\
&=&Q^{-3}n^{-3/2}\cdot\frac{2Q^{3/2}\sqrt{2}}{\sqrt{\pi}}\cdot\frac{2\sqrt{2}}{\sqrt{\pi}}(KQ+1)^{3/2}\\
&=&\frac{8}{\pi}\left(\frac{K+1/Q}{n}\right)^{3/2},\end{array}$$
which may be substituted into (\ref{berry_esseen}) to give
\begin{equation}\label{converge1}
\sup_x|F_n(x)-\Phi(x)|\le\frac{8}{\pi\sqrt{n}}.
\end{equation}
Now suppose $p\in\Lambda$ and consider the random variable
$$Y_i^n:=\frac{1}{Q\sqrt{n}}\left[X_{ip}\left(\sum_{j\in\Lambda}X_{ij}+z_i\right)-Q\right].$$
Note that we again have that the $\{Y^n_i\}$ are independent for $i=1,\ldots,n$, and $\EE Y_i=0$ and $\EE Y_i^2=\frac{K+1+1/Q}{n}<\infty$. Then the Berry-Esseen theorem guarantees that the distribution function, $F_n(x)$, of
$$\frac{\displaystyle\sum_{i=1}^n Y_i}{\sqrt{n}\,\EE Y_i^2}=\frac{\displaystyle\sum_{i=1}^n X_{ip}\left(\sum_{j\in\Lambda}X_{ij}+z_i-Q\right)}{nQ\left(\frac{K+1+1/Q}{n}\right)^{1/2}}$$
satisfies
\begin{equation}\label{berry_esseen2}
\sup_x|F_n(x)-\Phi(x)|\le\frac{\EE |Y_i^n|^3}{\sqrt{n}\left(\frac{K+1+1/Q}{n}\right)^{3/2}},
\end{equation}
and it remains to calculate $\EE|Y_i^n|^3$. Writing $Y_i^n=U^n_i+V^n_i$ where
$$U_i^n:=\frac{1}{Q\sqrt{n}}\left(X_{ip}^2-Q\right),\;\;\;\;V^n_i:=\frac{1}{Q\sqrt{n}}\left[X_{ip}\left(\sum_{j\in\Lambda\setminus p}X_{ij}+z_i\right)\right],$$
we have
\begin{equation}\label{max_bound}
\EE|Y^n_i|^3=\EE|U_i^n+V_i^n|^3\le\EE\left[2\max(|U_i^n|,|V^n_i|)\right]^3=8\max\left(\EE|U^n_i|^3,\EE|V_i^n|^3\right).
\end{equation}
We have 
$$U_i^n\sim\frac{1}{n}\left(\chi_1^2-1\right),$$
where $\chi_1^2$ denotes the chi-squared distribution with one degree of freedom, and calculation yields
$$\EE|U_i^n|^3=\frac{8}{n^4}\left[3+\frac{3}{\sqrt{2\pi}}+\sqrt{e}\left(1-2\,\textrm{erf}\frac{1}{\sqrt{2}}\right)\right].$$
Meanwhile, using a similar argument as in the $p\in\Lambda^C$ case yields
$$\EE|V_i^n|^3=\frac{8}{\pi}\left(\frac{K-1+1/Q}{n}\right)^{3/2}.$$
For sufficiently large $K$ and $n$ -- for example if $K\geq 2$ and $n\geq 12$ -- it is easy to show that $\EE|U^n_i|^3<\EE|V^n_i|^3$, which combines with (\ref{max_bound}) to give
$$\EE|Y_i|^3\le \frac{64}{\pi}\left(\frac{K-1+1/Q}{n}\right)^{3/2}\le\frac{64}{\pi}\left(\frac{K+1+1/Q}{n}\right)^{3/2},$$
which in turn combines with (\ref{berry_esseen2}) to give
\begin{equation}\label{converge2}
\sup_x|F_n(x)-\Phi(x)|\le\frac{64}{\pi\sqrt{n}}
\end{equation}
and (\ref{converge1}) and (\ref{converge2}) together establish the result.\hfill$\Box$\\
\\
We next give details of how Theorem~\ref{central_limit} can be used to obtain an accurate asymptotic numerical approximation for the performance of OST. Given a threshold $\lambda$, let $P_{\lambda}$ and $Q_{\lambda}$ be the number of $|g_i|$ above $\lambda$ from $\Lambda$ and $\Lambda^C$ respectively, divided by $\mC$. Then Theorem~\ref{central_limit} gives the following corollary.
\begin{corollary}
\begin{equation}\label{P_result}\mathbb{E}P_{\lambda}\rightarrow\delta\left[1-\Phi\left(\frac{\lambda-1}{\rho}\right)\right],
\end{equation}
\begin{equation}\label{Q_result}\mathbb{E}Q_{\lambda}\rightarrow(1-\delta)\left[1-\Phi\left(\frac{\lambda}{\rho}\right)\right],
\end{equation}
where $\Phi$ denotes the distribution function of the standard Normal distribution.
\end{corollary}
\textbf{Proof:} 
$$\begin{array}{rcl}\mathbb{E}P_{\lambda}&=&\displaystyle\sum_{i\in\Lambda}\displaystyle\frac{1}{\mC}\cdot\PP(G_i>\lambda)\\
&=&\displaystyle\frac{K}{\mC}\PP(G_{i^{\ast}}>\lambda),\end{array}$$
where $i^{\ast}\in\Lambda$, from which (\ref{P_result}) follows by taking limits and invoking Theorem~\ref{central_limit}. Similarly,
$$\begin{array}{rcl}\mathbb{E}Q_{\lambda}&=&\displaystyle\sum_{i\in\Lambda^C}\displaystyle\frac{1}{\CC}\cdot\PP(G_i>\lambda)\\
&=&\displaystyle\frac{\mC-K}{\mC}\PP(G_{\hat{i}}>\lambda),\end{array}$$
where $\hat{i}\in\Lambda^C$, from which (\ref{Q_result}) follows by taking limits and invoking Theorem~\ref{central_limit}.\hfill$\Box$\\
\\
Fixing an acceptance probability $\epsilon$, we solve
$$\mathbb{E}P_{\lambda}=(1-\epsilon)\delta,\;\;\;\;\mathbb{E}(P_{\lambda}+Q_{\lambda})=\delta,$$
to obtain a solution for $\lambda$ and $\rho$. Hence, asymptotically,
$$\rho=c(\delta,\epsilon),$$
where $c(\delta,\epsilon)$ is a constant which depends only upon $\delta$ and $\epsilon$. We therefore have, asymptotically, that
$$\frac{K+1/Q}{n}=c(\delta,\epsilon),$$
which is what was claimed in Section~\ref{theory}.

\bibliographystyle{unsrt}
\bibliography{chirp_bib}

\end{document}